\documentclass[a4paper,12pt]{article}
\usepackage{amssymb,textcomp,amsmath,graphicx,bm,subfig,comment,hyperref,float,color,cite}
\usepackage[perpage]{footmisc}

\long\def\symbolfootnote[#1]#2{\begingroup
	\def\thefootnote{\fnsymbol{footnote}}
	\footnote[#1]{#2}\endgroup}

\begin{document}

\pagestyle{empty}
\vspace*{1.5cm}

\begin{center}
  {\Large \bf Overshooting, Critical Higgs Inflation and Second Order
    Gravitational Wave Signatures} \\
\vspace{1cm}
{\large Manuel Drees\symbolfootnote[0]{$^{*}$drees@th.physik.uni-bonn.de}$^*$, Yong Xu\symbolfootnote[0]{$^{\dagger}$yongxu@th.physik.uni-bonn.de}$^\dagger$} \\
\vspace*{6mm}
{\it Bethe Center for Theoretical Physics and Physikalisches
Institut, Universit\"at Bonn,\\Nussallee~12, D-53115 Bonn, Germany} 
\end{center}

\begin{abstract}
  The self coupling $\lambda$ of the Higgs boson in the Standard Model
  may show critical behavior, i.e. the Higgs potential may have a
  point at an energy scale $\sim 10^{17-18}$ GeV where both the first
  and second derivatives (almost) vanish. In this case the Higgs boson
  can serve as inflaton even if its nonminimal coupling to the
  curvature scalar is only ${\cal O}(10)$, thereby alleviating
  concerns about the perturbative unitarity of the theory. We find
  that just before the Higgs as inflaton enters the flat region of the
  potential the usual slow--roll conditions are violated. This leads
  to ``overshooting'' behavior, which in turn strongly enhances scalar
  curvature perturbations because of the excitation of entropic
  (non--adiabatic) perturbations. For appropriate choice of the free
  parameters these large perturbations occur at length scales relevant
  for the formation of primordial black holes. Even if these
  perturbations are not quite large enough to trigger copious black
  hole formation, they source second order tensor perturbations, i.e.
  primordial gravitational waves; the corresponding energy density can
  be detected by the proposed space-based gravitational wave detectors
  DECIGO and BBO.

\end{abstract}
\clearpage
\setcounter{page}{1}
\pagestyle{plain}
\tableofcontents
\section{Introduction}

Inflation is a beautiful paradigm for the evolution of the very early
universe: it not only solves the problems of standard cosmology
\cite{Starobinsky:1980te,Guth:1980zm,Linde:1981mu}, but also generates the initial seeds for the formation of large structures via quantum fluctuations. The simplest inflationary models feature a single scalar field that slowly
``rolls down'' a rather flat potential (``slow--roll'' inflation).
The energy density during inflation is then dominated by the
potential, which leads to an approximately exponential expansion of
the universe. Often a separate ``inflaton'' field is introduced for
this purpose, but it would obviously be more economical to instead use
the single scalar Higgs field $\Phi$ of the Standard Model (SM) of
particle physics as inflaton. At TeV energies the Higgs self coupling
$\lambda$ is ${\cal O}(0.1)$; a coupling of this size leads to a
rather steep potential, which needs to be ``flattened'' by a large
non--minimal coupling to the Ricci scalar, $\xi \Phi^\dagger \Phi R$
\cite{Bezrukov:2007ep}. This yields \cite{Bezrukov:2007ep, Barvinsky:2008ia,Bezrukov:2008ej,Bezrukov:2008ut, Bezrukov:2009db,Barvinsky:2009fy,Barvinsky:2009ii}
$n_s \simeq 0.97, r \simeq 0.0034$ in agreement with observation, but
for a coupling $\xi \sim 10^{4}$.\footnote{According to
  \cite{Tenkanen:2019jiq} Higgs inflation is possible in Palatini
  gravity even without nonminimal coupling to $R$.} Such a large
coupling may violate perturbative unitarity cutoff \cite{Lerner:2009na,
  Hertzberg:2010dc, Burgess:2010zq}. 

This unitarity issue has been discussed at
length in the literature \cite{Bezrukov:2010jz,
  Allison:2013uaa,Calmet:2013hia, Burgess:2014lza,
  Bezrukov:2014ipa,Fumagalli:2016lls, Enckell:2016xse,
  Escriva:2016cwl,
  DeCross:2015uza,Ema:2016dny,Sfakianakis:2018lzf}. In particular, it is argued in \cite{Bezrukov:2010jz} that  the unitarity bound could be higher than the inflationary scale if the inflaton field background is taken into account in the effective field theory (EFT) setup.  However it is shown that unitarity would still be violated due to the violent production of longitudinal  gauge bosons during preheating, where modes with momenta (much) higher than the perturbative cutoff (even with the background field) can be excited \cite{DeCross:2015uza,Ema:2016dny,Sfakianakis:2018lzf} \footnote{We thank the anonymous referee for bringing these references into our attention.}. So unitarity would  still be an issue in the post-inflationary phase even though it is not during inflation. This has motivated  extensions of original scenario to some UV completed theory  by considering Higgs inflation with additional field(s) beyond the SM \cite{Giudice:2010ka,Kawai:2014doa,Haba:2014zda,Ko:2014eia, Haba:2014zja, He:2014ora,Kahlhoefer:2015jma,Ballesteros:2016xej}; however, these models lack the simplicity of the original suggestion.\footnote{The unitarity problem could also be resolved by considering the new Higgs inflation scenario \cite{Germani:2010gm,Escriva:2016cwl,Fumagalli:2017cdo}.}

On the other hand, at the large field values where Higgs inflation may
have occurred, the value of $\lambda$ is expected to be quite
different than at the weak scale. This difference is described by
renormalization group equations (RGE). At the one--loop order
$\lambda$ is driven to larger values by Higgs self--interactions
(i.e. the $\lambda^2$ term contributes with positive sign in the RGE)
and by electroweak gauge interactions, but is reduced by Yukawa
interactions, the by far most important one being that of the top. The
evolution of the top Yukawa coupling in turn is also affected by QCD
interactions. For the measured value of the mass of the physical Higgs
boson (which determines $\lambda$ at the electroweak energy scale),
the two--loop RGE indicate that $\lambda$ may show {\em critical
  behavior} at energy scale $\sim 10^{17-18}$ GeV, i.e. $\lambda$ and
its first derivative, described by its beta function, can both be very
small \cite{Buttazzo:2013uya}.\footnote{For the current central values
  of the top mass (which determines the top Yukawa coupling at the
  weak scale) and the gauge couplings, within the SM $\lambda$ reaches
  zero already near $10^{11}$ or $10^{12}$ GeV. However, the
  interpretation of the experimentally measured top mass is somewhat
  uncertain \cite{Hoang:2020iah}. Moreover, at very high energy scales
  new degrees of freedom may appear. For example, new strongly
  interacting particles without direct coupling to the Higgs boson
  will increase the strong coupling, and thereby reduce the top Yukawa
  coupling, which in turn increases $\lambda$, at energies above the
  masses of these particles.} This implies that the potential becomes
very flat in this region \cite{Hamada:2013mya}, which can give rise to
``critical Higgs inflation'' (CHI).\footnote{For only slightly
  different values of the relevant parameters the Higgs potential can
  also have a second minimum at these large field values. This might
  not lead to a successful model of inflation since the
  Higgs field might get ``stuck'' in this minimum, in which
  case inflation would not end. However if the inflaton carries sufficient  kinetic energy, it can still climb uphill and reach to the end of inflation. See  Refs. \cite{Jinno:2017jxc,Jinno:2017lun} for Hillclimbing inflationary scenarios. } Since $\lambda$ is small, one only needs a non--minimal coupling $\xi \sim \mathcal{O}(10)$\cite{Bezrukov:2014bra, Hamada:2014wna, Hamada:2014iga}; see also
\cite{Salvio:2017oyf, Masina:2018ejw, Bezrukov:2017dyv, Salvio:2018rv}
for recent investigations concerning CHI and \cite{Rubio:2018ogq} for
a comprehensive review of Higgs inflation.

In addition to reproducing the measured CMB power spectrum accurately,
recently some other cosmological implications of the CHI scenario have
been investigated. In particular ref.~\cite{Ezquiaga:2017fvi} showed
that curvature perturbations are greatly enhanced at the length scales
that leave the horizon when the inflaton field enters the very flat
region of the potential; this might even lead to the formation of a
cosmologically significant abundance of primordial black holes
(PBH).\footnote{See also \cite{Garcia-Bellido:2017mdw,Gong:2017qlj,Germani:2017bcs, Gao:2018pvq, Dalianis:2018frf,Ezquiaga:2018gbw,Biagetti:2018pjj, Ozsoy:2018flq,
    Cicoli:2018asa,Hertzberg:2017dkh,
    Ballesteros:2017fsr,Motohashi:2017kbs, Kannike:2017bxn,
    Inomata:2017okj,Cheng:2018qof,Fu:2019ttf} for more recent similar works where
  PBHs are produced through large quantum fluctuations when the
  inflaton enters a very flat stretch of the potential.} In fact, in
the simplest approximation the strength of the density perturbations
is inversely proportional to the first derivative $V'$ of the inflaton
potential $V$. It is thus tempting to associate a spike in the
spectrum of density perturbations with an ``ultra--slow roll'' (USR)
phase in which the inflaton field moves extremely slowly because it
traverses a very flat piece of the potential. We will see in
Section~\ref{adandnonad} that this is not the whole story: the largest
enhancement actually does not happen in the USR phase, but during a
transitionary ``overshooting'' stage just before USR where the
inflaton potential has a sizable curvature, so that the slow--roll
(SR) conditions are violated. We will show that in this case curvature
perturbations can continue to grow even after they cross out of the
horizon, since the perturbations are no longer adiabatic,
i.e. ``entropic'' perturbations are excited. To our knowledge, this is
the first investigation of significant effects due entropic
perturbations on observable inhomogeneities.

The enhanced scalar curvature perturbations are expected to source
tensor perturbations at second order, as investigated in recent papers
\cite{Gong:2017qlj,Bartolo:2018evs,Hajkarim:2019nbx,Liu:2020oqe,Xu:2019bdp,Fu:2019vqc,Bartolo:2018rku,Clesse:2018ogk,	Kohri:2018awv,Braglia:2020eai}. In this paper, we improve and further
extend the analysis in \cite{Ezquiaga:2017fvi} in several ways.  First
of all, the power spectrum in \cite{Ezquiaga:2017fvi} is calculated
within the SR approximation at all scales; however this assumption
does not hold during the overshooting phase. We therefore use the more
accurate numerical Mukhanov--Sasaki formalism. Moreover, detailed
explanations for our numerical results for the power spectrum are
given; this might help to understand features of the power spectrum in
other PBH production scenarios from single field inflation with a
near--inflection point, for example \cite{Gong:2017qlj, Cheng:2018qof,
  Ezquiaga:2018gbw}. With a more realistic result of scalar power
spectrum, we investigate the second order gravitational wave (GW)
signatures arising from large scalar curvature perturbations in the
CHI scenario. We show that such signatures can be detected by several
proposed space based GW experiments. The calculation of the PBH
density is fraught with considerable uncertainty
\cite{Germani:2018jgr,Harada:2013epa}. Our result indicates that, at
least for CHI inflation, an inflationary GW signal should be
detectable in all cases that could conceivably lead to sizable PBH
production, i.e. a failure to detect the latter in future experiments
would exclude the possibility that PBHs contribute significantly to
cosmological dark matter.

The reminder of this paper is organized as follows. We first revisit
the curvature perturbations under adiabatic and non--adiabatic
conditions in Section~\ref{adandnonad}; in particular, we show that
for non--adiabatic conditions the amplitude of curvature perturbations
does not necessarily remain constant after horizon crossing, in
contrast to the usual SR treatment. In
Section~\ref{secchi} we set up the CHI scenario. Using the general
Mukhanov--Sasaki formalism to compute the power spectrum, we show that
the standard SR approximation to calculate the power spectrum fails
when the inflaton enters an overshooting phase, even if we use
``Hubble'' rather than ``potential'' SR parameters. In
Section~\ref{gwsig} we discuss second order GW signatures induced by
the scalar curvature perturbations. Finally we summarize our results in
Section~\ref{conclusion}. In appendix \ref{solums} we give a quick
review of the Mukhanov--Sasaki equation and its analytical solution
for a (quasi) de Sitter spacetime.

\section{Evolution of Curvature Perturbations}
\label{adandnonad}

\subsection{Adiabatic and Entropic Perturbations}

In single field SR inflation the quantum fluctuations are {\em
  adiabatic}. As a result the perturbations of
all inflaton field dependent quantities $X_i$ share the same phase
trajectory \cite{Riotto:2002yw}:
\begin{equation}
\frac{\delta X_i(t, \bm{x}) }{\dot{\bar{X}}_i} = \frac{\delta X_j(t, \bm{x}) }
{\dot{\bar{X}}_j} = ... = \delta t(\bm{x}),
\end{equation}
where $i,j$ denote different observables, $\bar{X}_i$ represents the
average (background) of $X_i$, $\delta X_i \equiv X_i - \bar X_i$ its
perturbation, and $\dot{\bar{X}}$ the time derivative. In particular,
using $X_1 = p$ (pressure) and $X_2 = \rho$ (energy density),
adiabatic perturbations satisfy
\begin{equation}
\frac {\delta p_{\rm ad}} { \dot{\bar{p}} }  =
\frac {\delta \rho_{\rm ad} } { \dot{\bar{\rho}} }
\Rightarrow \delta p_{\rm ad} \equiv
\frac { \dot{\bar{p}}} { \dot{\bar{\rho}} } \delta \rho_{\rm ad}.
\end{equation}
Energy density and pressure are defined via the energy--momentum
tensor $T_\mu^\nu$, with $\rho = T_0^0$ and $p = -\sum_{i=1}^3 T_i^i/3$.

However, in some cases the perturbation may not be adiabatic, for
example when there are multiple fields interacting with the inflaton
\cite{Gordon:2000hv} or while the universe undergoes a non--SR
inflationary phase, see Sec.~\ref{srusr}. Thus more generally the
pressure perturbation can be decomposed into an adiabatic part and an
entropic (i.e. non--adiabatic) one:
\begin{equation}
\delta p = \delta p_{\rm ad} + \delta p_{\rm en},
\end{equation}
i.e. $ \delta p_{\rm en} \equiv \delta p -\delta p_{\rm ad}$. We will
show in the next subsection that this distinction has significant
impact on the evolution of curvature perturbations.

\subsection{Evolution of Curvature Perturbations in SR, USR and
	Overshooting Phases}
\label{srusr}

In order to relate inflationary predictions and cosmic microwave
background (CMB) measurements the gauge invariant scalar quantity
called curvature perturbation is usually introduced; it is defined by
\cite{Bardeen:1980kt, Bardeen:1983qw}
\begin{equation} \label{gaugeinvariant}
-\zeta(t,\bm{x}) \equiv \Psi(t,\bm{x}) +
\frac {H} { \dot{\bar{\rho}}(t) } \delta{\rho}(t,\bm{x})\,.
\end{equation}
Here $H$ denotes the Hubble parameter, and $\Psi(t,\mathbf{x})$ is a
scalar function of coordinates. Physically $\zeta$ represents the
spatial curvature of hypersurfaces with uniform energy density
\cite{Baumann:2009ds}. Since the power of the two--point correlation
function of $\zeta$ is related to the CMB temperature anisotropies,
one has to compute the power spectrum of $\zeta$ for a given
inflationary model. This is usually done in Fourier space, where
$\zeta(t,\bm{k})$ is associated to perturbations at a comoving length
scale $1/k$ with $k=|\bm{k}|$. As we will review below, under SR
conditions the power spectrum can be computed when some mode $k$
crosses the horizon, since $\zeta$ is frozen at super--horizon scale,
i.e. it remains constant once $k \ll aH$ where $a$ is the
(dimensionless) scale factor in the Friedman--Robertson--Walker (FRW)
metric. However, whenever the universe deviates from SR expansion, we
must in general take the super--horizon evolution of $\zeta$ into
account; unfortunately this considerably complicates the accurate
computation of the power spectrum.

Using energy-momentum conservation it can be shown that the evolution of
$\zeta$ is given by \cite{Baumann:2009ds}
\begin{equation} \label{superhorizon}
\dot{\zeta} = -H \frac{\delta p_{\rm en}}{\bar{\rho} + \bar{p}} -\Pi\,.
\end{equation}
Here $\delta p_{\rm en}$ is the non--adiabatic component of the
pressure perturbation, and $\Pi$ is defined as
\begin{equation} \label{superhorizon2}
\frac{\Pi}{H} = -\frac{k^2}{3a^2H^2} \left[ \zeta -
\Psi_B \left( 1 - \frac{2 \bar{\rho}} {9(\bar{\rho} + \bar{p} )}
\frac {k^2} {a^2H^2} \right) \right] \,,
\end{equation}
where $\Psi_B$ is a Bardeen potential \cite{Bardeen:1980kt} which does
not depend on $k$. We thus see that at super horizon scales, i.e. for
$k \ll aH$, the second term in eq.(\ref{superhorizon}) can be
neglected. If in addition the perturbations are adiabatic, i.e. if
$\delta p_{\rm en}$ can be neglected, then $\zeta$ is conserved on
super--horizon scales. Weinberg showed~\cite{Weinberg:2003sw} that
solutions with $\delta p_{\rm en} = 0$, and hence
$\dot{\zeta} \rightarrow 0$ for $k \ll aH$, always exist. We will see
that this is the only solution if the universe follows a SR expansion,
i.e. for a quasi--de Sitter spacetime; however, under overshooting
conditions a non--adiabatic solution also exists, and can lead to
large enhancement of the curvature perturbation.

Since we define $\delta X = X(\phi) - X(\bar\phi)$, $\delta X$ can
include terms that are of higher order in the field perturbation
$\delta \phi$.\footnote{Eq.(\ref{superhorizon}) holds to linear order
  in the perturbation $\delta p_{\rm en}$. However, this does not
  imply that $\delta p_{\rm en}$ is dominated by terms that are linear in
  $\delta \phi$.} We assume that only a single (inflaton) field $\phi$
has sizable perturbations. Moreover, we are interested only in
super--horizon modes where all gradient terms can be neglected. The
energy density and pressure are thus given by
\begin{equation} \label{rhop}
\rho = \frac{1}{2} \left( \dot{\phi} \right)^2 + V(\phi)\,; \ \ \ \
p = \frac{1}{2} \left( \dot{\phi} \right)^2 - V(\phi)\,.
\end{equation}
The same equations also describe $\bar \rho$ and $\bar p$, which
$\phi \rightarrow \bar \phi$ on the right--hand side. Super--horizon
size non--adiabatic pressure perturbations are then given by:
\begin{equation} \label{adnonad}
\begin{split}
\delta p_{\rm en} &= \delta p- \frac { \dot{\bar{p}} } { \dot{\bar{\rho}} }
\delta \rho\\
& =  \left[ \left( \frac{1}{2} \dot{\phi}^2 - V(\phi)\right)
- \left( \frac{1}{2} (\dot{\bar{\phi}})^2 - V(\bar{\phi}) \right) \right] \\
&- \frac{ \dot{\bar{\phi}} \ddot{\bar{\phi}} - \dot{\bar{\phi}}
	V^{\prime}_{\bar{\phi}} } { \dot{\bar{\phi}} \ddot{\bar{\phi}} +
	\dot{\bar{\phi}} V^{\prime}_{\bar{\phi}} }  \left[ \left(
\frac{1}{2} \dot{\phi}^2 + V(\phi) \right) - \left(
\frac{1}{2} (\dot{\bar{\phi}})^2 + V(\bar{\phi})\right)\right] \\
&= \left[ \frac{1}{2} \left( \dot{\bar{\phi}} +\delta \dot{\phi} \right)^2
- V(\bar{\phi} + \delta \phi ) - \left( \frac{1}{2}(\dot{\bar{\phi}})^2
- V(\bar{\phi}) \right) \right] \\
&- \frac{ \ddot{\bar{\phi}} - V^{\prime}_{\bar{\phi}} }
{ \ddot{\bar{\phi}} + V^{\prime}_{\bar{\phi}} }  \left[ \frac{1}{2}
\left( \dot{\bar{\phi}} + \delta \dot{\phi} \right)^2
+ V(\bar{\phi} + \delta \phi) - \left( \frac{1}{2} (\dot{\bar{\phi}})^2
+ V(\bar{\phi}) \right) \right]\,.
\end{split}
\end{equation}
Using Taylor expansion for the potential up to second
order\footnote{The second order of the perturbation gives the
  two--point correlation function; higher orders contribute to
  non--Gaussian corrections to the power spectrum of the perturbation,
  which are beyond the scope of this paper.} of $\delta \phi$ , we
obtain
$V(\bar{\phi} + \delta \phi) = V(\bar{\phi}) + V^{\prime}_{\bar{\phi}}
\delta \phi +\frac{1}{2}V^{\prime \prime}_{\bar{\phi}}(\delta
\phi)^2$, where $V^{\prime}_{\bar{\phi}}$ denotes
$dV(\bar{\phi})/d\bar{\phi}$. Using this expansion,
eq.(\ref{adnonad}) becomes
\begin{equation} \label{adnonad2}
\begin{split}
\delta p_{\rm en} &= \left( \dot{\bar{\phi}} \delta \dot{\phi} +
\frac{1}{2} (\delta \dot{\phi})^2 - V^{\prime}_{\bar{\phi}} \delta \phi
-\frac{1}{2} V^{\prime \prime}_{\bar{\phi}} (\delta \phi)^2 \right) \\
&- \frac{ \ddot{\bar{\phi}} - V^{\prime}_{\bar{\phi}} }
{ \ddot{\bar{\phi}} + V^{\prime}_{\bar{\phi}} } \left(
\dot{\bar{\phi}} \delta \dot{\phi} + \frac{1}{2} ( \delta \dot{\phi})^2
+ V^{\prime}_{\bar{\phi}} \delta \phi + \frac{1}{2} V^{\prime \prime}_{\bar{\phi}}
(\delta \phi)^2 \right) \,.
\end{split}
\end{equation}
For a strict de Sitter spacetime $V$ has to be constant, i.e. all
derivatives of $V$ vanish. It is easy to see that $\delta p_{\rm en} = 0$
in this case. However, during realistic SR inflation,
$|\ddot{\bar{\phi}}| \ll \left| V^{\prime}_{\bar{\phi}} \right|$, so
that eq.(\ref{adnonad2}) reduces to
\begin{equation} \label{adnonad3}
\begin{split}
\delta p_{\rm en} &= \left( \dot{\bar{\phi}} \delta \dot{\phi}
+ \frac{1}{2} (\delta \dot{\phi})^2 - V^{\prime}_{\bar{\phi}} \delta \phi
- \frac{1}{2} V^{\prime \prime}_{\bar{\phi}} (\delta \phi)^2 \right) \\
&- (-)  \left( \dot{\bar{\phi}} \delta \dot{\phi} + \frac{1}{2}
( \delta \dot{\phi} )^2 + V^{\prime}_{\bar{\phi}} \delta \phi
+\frac{1}{2} V^{\prime \prime}_{\bar{\phi}} (\delta \phi)^2 \right)\\
&=2 \dot{\bar{\phi}} \delta \dot{\phi} + (\delta \dot{\phi})^2\,.
\end{split}
\end{equation}

In order to see that $\delta p_{\rm en}$ is indeed very small during
SR, consider the equation of motion for $\delta \phi$
\cite{Riotto:2002yw,Espinosa:2017sgp}:
\begin{equation} \label{eomquanfluc}
\delta \ddot{\phi} + 3 H\delta\dot{\phi} - \frac{\nabla^2 \delta \phi}{a^2}
+ V^{\prime \prime} \delta \phi = 0\,.
\end{equation}
In momentum space this becomes
\begin{equation} \label{momentumspace}
\delta \ddot{\phi}_k + 3 H \delta \dot{\phi}_k + \frac{k^2}{a^2}\delta \phi_k
+ V^{\prime \prime} \delta \phi_k = 0\,.
\end{equation}
In analyses of inflationary dynamics it is often useful to trade the
time for the number of e--folds $N$ via $dN = Hdt$;
eq.(\ref{momentumspace}) then becomes
\begin{equation} \label{momentumspace2}
\frac{d^2 \delta \phi_k }{dN^2} + 3 \frac {d \delta \phi_k } {dN}
+ \frac {k^2} {a^2H^2} \delta \phi_k
+ \frac {V^{\prime \prime}} {H^2} \delta \phi_k = 0\,.
\end{equation}
At super--horizon scales ($k\ll a H$) the third term in
eq.(\ref{momentumspace2}) can be neglected. Moreover, during SR the
total energy density is dominated by the potential energy, so
that\footnote{We set the reduced Planck scale
	$M_P \simeq 2.4\cdot10^{18}$ GeV to $1$ in the following.}
$H^2 \simeq \frac{1}{3}V$. Finally, we introduce the second potential SR
parameter $\eta_V \equiv \frac{V^{\prime \prime}}{V}$, which has to be
small during SR inflation. Eq.(\ref{momentumspace2}) can then be
written as:
\begin{equation} \label{momentumspace4}
\frac{d^2 \delta \phi_k }{dN^2} + 3 \frac{d \delta \phi_k }{dN}
+ 3 \eta_V \delta \phi_k = 0\,.
\end{equation}

For constant $\eta_V$ with $|\eta_V| \ll 1$ the solution of
eq.(\ref{momentumspace4}) is given by
\begin{equation} \label{sol1}
\delta \phi_k \simeq C_1 {\rm e}^{-3N} + C_2 {\rm e}^{-\eta_V N}\,,
\end{equation}
where the constants $C_{1,2}$ are of order $H/(2\pi)$, which
determines the size of $|\delta \phi|$ due to quantum fluctuations
during SR inflation. This implies
$\frac{d \delta \phi_k }{dN} \leq H \left[ {\rm e}^{-3N} + {\cal
	O}(\eta_V) \right]$, or equivalently
$\delta \dot{\phi} \leq H^2 \left[ {\rm e}^{-3Ht} + {\cal O}(\eta_V)
\right]$. Moreover, during SR
$|\dot{\bar{\phi}}| \simeq \frac {|V^{\prime}|}{3H} = H \frac{
	|V^{\prime}| } {V} = H \sqrt{2 \epsilon_V}$, where
$\epsilon_V = \frac{1}{2} \left( \frac{V^{\prime}}{V}\right)^2$
denotes the first potential SR parameter. Thus we see
$\dot{\bar{\phi}}$ is also rather small and nearly constant.

Inserting these estimates in eq.(\ref{adnonad3}) and using
eqs.(\ref{superhorizon}) and (\ref{rhop}) we find for the time evolution
of the curvature perturbations at super--horizon scales:
\begin{equation} \label{zetadot2}
\begin{split}
|\dot{\zeta}| &= 2H \left[ \left| \frac {\delta \dot{\phi}}
{ \dot{\bar{\phi}} } \right| + \frac{1}{2} \left(
\frac {\delta \dot{\phi}} { \dot{\bar{\phi}} } \right)^2 \right] \\
&\leq  2H \frac{3H^3 \left[ {\rm e}^{-3Ht} + {\cal O}(\eta_V) \right] }
{|V^{\prime}|}  + H \left( \frac { 3H^3 \left[ {\rm e}^{-3Ht} + {\cal O}(\eta_V)
	\right]} {V^{\prime}} \right)^2 \\
&\simeq  2H^2 \frac { V \left[ {\rm e}^{-3Ht} + {\cal O}(\eta_V) \right] }
{|V^{\prime}|} + H^3 \left( \frac{ V \left[ {\rm e}^{-3Ht} + {\cal O}(\eta_V)
	\right] } {V^{\prime}} \right)^2 \\
& = \frac{2H^2}{\sqrt{2 \epsilon_V }} \left[ {\rm e}^{-3Ht} + {\cal O}(\eta_V)
\right] +  \frac{H^3}{2 \epsilon_V } \left[ {\rm e}^{-3Ht} + {\cal O}(\eta_V)
\right]^2\,.
\end{split}
\end{equation}
Note that $\frac{H}{\epsilon_V} \sim \sqrt{\mathcal{P}_\zeta} \sim |\zeta|$,
where $\mathcal{P}_\zeta$ is the power in the perturbation. At the length scales
probed by the CMB, $\mathcal{P}_\zeta \sim 10^{-9}$ is very small. More
importantly, the last line in eq.(\ref{zetadot2}) shows that the time
derivative of $\zeta$ is suppressed by the SR parameter $\eta_V$ once
the exponentially decaying part of $\delta \dot \phi_k$ can be
ignored. This completes our argument that in the SR regime $\zeta$ is
(nearly) constant once the mode crosses out of the horizon.

However, whenever the universe deviates from SR expansion,
$\dot{\zeta}$ may no longer be negligible even in the super--horizon
regime due to the entropic pressure perturbation. Solutions of this
kind correspond to what Weinberg called the {\em non-adiabatic} mode
\cite{Weinberg:2003sw}. Of special interest to us is the situation
where the acceleration term is much larger than the derivative of the
potential, i.e.  $|\ddot{\bar{\phi}}| \gg
|V^{\prime}_{\bar{\phi}}|$. In this case eq.(\ref{adnonad2}) becomes
\begin{equation} \label{adnonad4}
\begin{split}
\delta p_{\rm en} &= \left( \dot{\bar{\phi}} \delta \dot{\phi}
+ \frac{1}{2}(\delta \dot{\phi})^2 - V^{\prime}_{\bar{\phi}} \delta \phi
-\frac {1} {2} V^{\prime \prime}_{\bar{\phi}} (\delta \phi)^2 \right) \\
&- (+)  \left( \dot{\bar{\phi}} \delta \dot{\phi} + \frac {1} {2}
(\delta \dot{\phi})^2 + V^{\prime}_{\bar{\phi}} \delta \phi
+\frac {1} {2} V^{\prime \prime}_{\bar{\phi}}(\delta \phi)^2 \right)\\
&= -2\left( V^{\prime}_{\bar{\phi}} \delta \phi
+ \frac {1} {2} V^{\prime \prime}_{\bar{\phi}} (\delta \phi)^2 \right) \\
&\approx -V^{\prime \prime}_{\bar{\phi}} (\delta \phi)^2\,.
\end{split}
\end{equation}
In the last step we have neglected $V^{\prime}_{\bar{\phi}}$ relative to
$V^{\prime \prime}_{\bar{\phi}} \delta \phi$. 

Our assumption $|\ddot{\bar{\phi}}| \gg |V^{\prime}_{\bar{\phi}}|$ is
equivalent to having the second (Hubble) SR parameter
$\eta_H \equiv -\ddot{\bar{\phi}} / (H\dot{\bar{\phi}}) \approx 3$, 
which is manifestly not smaller than unity, i.e. the SR conditions are
violated. Some recent papers state that this scenario corresponds to
an USR phase. We disagree with this interpretation. The expression
``ultra--slow'' roll implies that the inflaton field evolves even more
slowly than during SR, which happens when the potential becomes very
flat, in which case the SR parameters should also be small. In other
words, the spacetime during USR should be even more de Sitter like
than that during SR, so the perturbations should be even more
adiabatic than during SR, and the evolution of $\zeta$ at
super--horizon scales should be even more suppressed. Hence the SR
approximation for the power spectrum, where it is computed at horizon
crossing, should work even better in a true USR phase, rather than
breaking down.

So the phase with $|\ddot{\bar{\phi}}| \gg |V^{\prime}_{\bar{\phi}}|$
cannot correspond to USR, but to an intermediate transition
``overshooting'' stage (also mentioned in \cite{Germani:2017bcs})
between SR to USR, where the curvature of the potential is sizable but
the first derivative is already rather small. We will see below that
critical Higgs inflation can indeed lead to a situation where
$\eta_H \simeq 3$ for several e--folds of inflation. 

In order to get a first qualitative understanding of such an ``overshooting''
stage, we insert the final result of eq.(\ref{adnonad4}) into
eq.(\ref{superhorizon}):
\begin{equation} \label{curvature_evolution}
\dot{\zeta}  \approx H V^{\prime \prime}_{\bar{\phi}}  \left(
\frac {\delta \phi} {\dot{\bar{\phi}}} \right)^2\,.
\end{equation}
Note that $\ddot{\bar{\phi}} + 3H\dot{\bar{\phi}} =0$ implies
$\dot{\bar{\phi}} \propto {\rm e}^{-3N}$ so that
$1/\dot{\bar\phi}^2 \propto {\rm e}^{6N}$. As a result the derivative
$d \zeta / dN$ grows exponentially during this overshooting region.
Since neither $\delta \phi$ nor the curvature $V^{\prime\prime}$ are
(approximately) constant during this overshooting epoch,
eq.(\ref{curvature_evolution}) is not so well suited for a
quantitative treatment of the evolution of the curvature perturbations;
this can be done using the Mukhanov--Sasaki equation, as will be
described in the next section. However, we can already conclude that
$|\ddot{\bar{\phi}}| \gg |V^{\prime}_{\bar{\phi}}|$ implies that the
curvature perturbation is {\em not} frozen at the super--horizon
scales, and even increases significantly if the potential has a large
positive curvature $V^{\prime \prime}$. Of course, the usual SR
treatment of approximating the final power spectrum by its value at
horizon crossing will then no longer work. We consider
eq.(\ref{curvature_evolution}) and its consequences to be one of the
central results of this paper, which is applicable whenever an
overshooting epoch occurs during the evolution of the inflaton
field. In the next section we will explore the quantitative
consequences for the case of critical Higgs inflation.

Before concluding this section we briefly discuss the evolution of the
perturbations after inflation ends. During matter domination the
pressure is by definition negligible. During radiation domination,
$p \simeq \rho/3$ holds {\em locally}, which again implies
$\delta p_{\rm en} = 0$. Hence curvature perturbations remain frozen on
super--horizon scales after inflation.

\section{Critical Higgs Inflation}
\label{secchi}

In this section we discuss critical Higgs inflation, with emphasis on
the enhancement of curvature perturbations associated with an overshooting
region. We first describe the basic set--up in the Jordan and Einstein frames.
In the second subsection we analyze the inflationary dynamics in the
Einstein frame and connect it to CMB observables. In sec.~3.3 we show that
SR conditions are violated in the overshooting region, just before
the inflaton enters the very flat part of the potential. We then review
the Mukhanov--Sasaki formalism, which we use in sec.~3.4 for a detailed
numerical investigation. 

\subsection{Formalism}

Starting point of the analysis is the action in the Jordan frame (in Planckian
units, where $M_p = \sqrt{\frac{1}{8\pi G}}=1$):
\begin{equation}
\begin{split}
S_J &= -\int d^4x \sqrt{-g} \left[ \frac{1}{2} \left(1 + \xi(h) h^2\right)R
- \frac {1} {2} \partial_\mu h \partial^\mu h
+ \frac {\lambda(h)} {4} h^4 \right] \\
&= -\int d^4x \sqrt{-g} \left[ f(h) R
- \frac {1} {2} \partial_\mu h \partial^\mu h
+ \frac {\lambda(h)} {4} h^4 \right]\,.
\end{split}
\end{equation}
In the second line we have introduced the function
$f(h) = \frac{1}{2} (1 + \xi h^2)$.  The crucial observation
\cite{Ezquiaga:2017fvi} is that for realistic values of the relevant
SM parameters, the running Higgs self coupling $\lambda$ attains a
minimum at scale $\mu$. Near this minimum it can then be expanded as:
\begin{equation}
\lambda(h) = \lambda_0 + b_{\lambda} \ln^2\left( \frac {h} {\mu} \right)\,.
\end{equation}
The running non--minimal coupling $\xi$ to the Ricci scalar is also
expanded around scale $\mu$:
\begin{equation}
\xi(h) = \xi_0 + b_\xi \ln\left( \frac {h}{\mu} \right)\, ;
\end{equation}
since $\xi$ does not have an extremum at scale $\mu$, the leading energy
dependence is described by a term linear, rather than quadratic, in
$\ln(h/\mu)$.

While the matter part of the Jordan frame action has its canonical
form, this is not true for the gravitational part, unless
$|\xi(h) h^2| \ll 1$.  In order to use standard results for the
inflationary dynamics we transform to the Einstein frame, where
gravity is described by the well--known Einstein--Hilbert action and the
inflaton is described by a canonically normalized field $\chi$. To that end
we first utilize a conformal transformation to the Einstein frame:
\begin{equation}
\tilde{g}_{\mu \nu} = \Omega^2 g_{\mu \nu} = ( 1 + \xi h^2) g_{\mu \nu}\,.
\end{equation}
Then we use a field redefinition to obtain a canonical kinetic term
\cite{Kaiser:2010ps}; it is defined by:
\begin{equation} \label{chih}
\begin{split}
\frac {d\chi} {dh} &= \sqrt{ \frac { f(h) + 3f(h)^{\prime \ 2} } {2f(h)^2}} \\
&= \sqrt { \frac { 1 + \xi h^2 + 6 \left( h\xi
		+ \frac{1}{2} h^2 \xi^{\prime}\right)^2 } { ( 1 + \xi h^2) ^2 } }\,.
\end{split}
\end{equation}
After these transformations the action becomes
\begin{equation}
S_E =- \int d^4x \sqrt{ -\tilde{g}} \left[ \frac{1}{2} \tilde{R}
- \frac{1}{2} \partial_\mu \chi \partial^\mu \chi + V(\chi) \right]\,.
\end{equation}
While the gravitational part as well as the kinetic energy term in the
action now have the standard form, the inflationary potential has
become more complicated:
\begin{equation}
V(\chi) = \frac {1} {\Omega(\chi)^4} \frac {\lambda(h(\chi))} {4} h(\chi)^4\,.
\end{equation}
It is convenient to introduce the quantities
\[
x = \frac{h}{\mu},\  a = \frac {b_{\lambda}} {\lambda_0}, \
b = \frac {b_{\xi}} {\xi_0}, \ c = \xi_0 \mu^2 \ \mathrm{and}\
V_0 = \frac{\lambda_0\mu^4}{4}\,.
\]
The inflaton potential can then be written as
\begin{equation}
V(x) = \frac{ V_0 (1 + a \ln^2x) x^4}
{ \left[ 1 + c ( 1 +b \ln x) x^2 \right]^2 }\,.
\label{potential_x}
\end{equation}
Note that for nonminimal coupling $\xi \neq 0$ the potential
approaches a constant as $x \rightarrow \infty$; it is this
``flattening'' which allows inflation. Consistency with the CMB
observables (see below) and with current measurements of SM parameters
can be obtained for parameter values in the ranges
\cite{Ezquiaga:2017fvi} $\lambda_0 \sim (0.01 - 8) \times 10^{-7}$,
$b_{\lambda} \sim (0.008 - 4) \times 10^{-6}$,
$\xi_0 \sim (0.5 - 15)$, $\mu^2 \sim (0.05 - 1.2)$ and\footnote{The
  large running of the non-minimal coupling $b_{\xi}$ could arise from
  the scalaron degree of freedom \cite{Ema:2019fdd,Salvio:2014soa}.}
$b_{\xi} \sim (1 - 18)$. In order to compare our calculations,
especially the power spectrum, with those in \cite{Ezquiaga:2017fvi}
based on the SR approximation, we mainly work with their
representative set of parameters :
\begin{equation} \label{modelparameter}
\lambda_0 = 2.23 \times 10^{-7}, \ \xi_0=7.55, \ \mu^2 = 0.102,
\ b_{\lambda} = 1.2\times10^{-6}, \ \mathrm{and} \ b_{\xi} = 11.5.
\end{equation}

\begin{figure}[htbp!]
\centering
\includegraphics[width=.6\paperwidth, keepaspectratio]{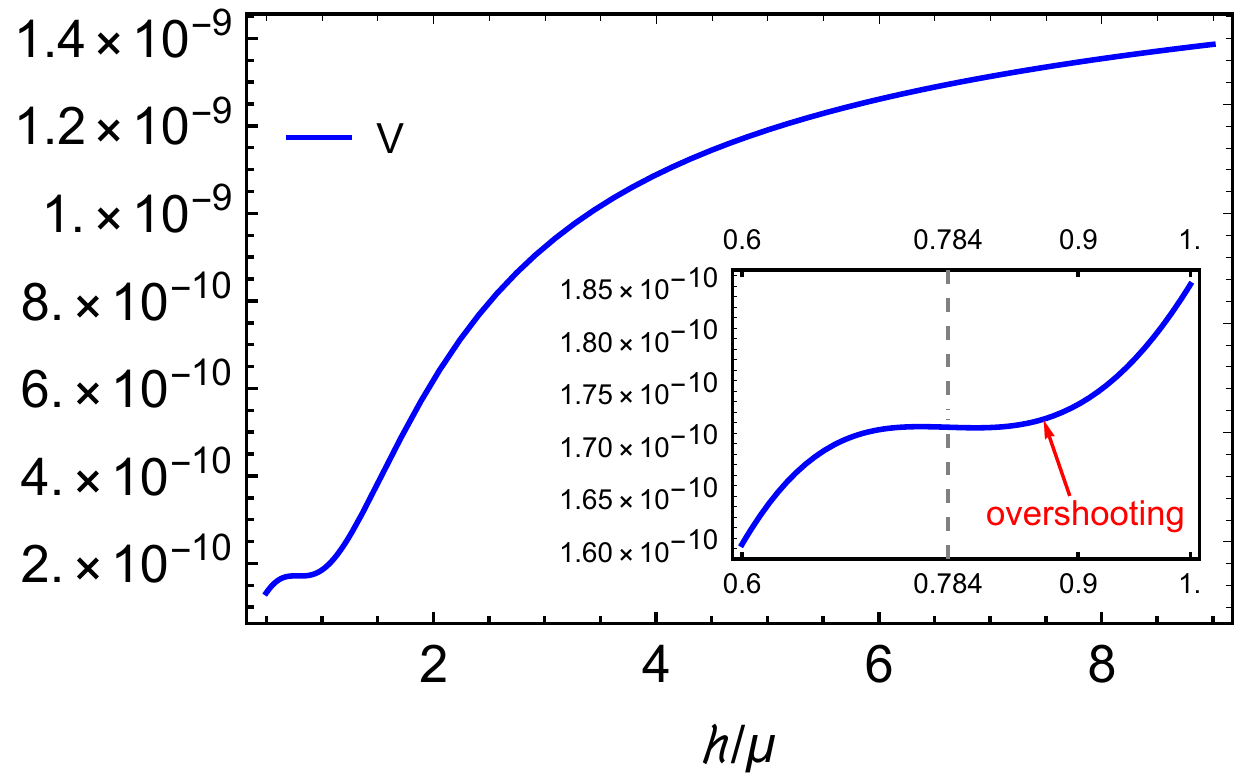}
\caption {Inflaton potential as function of the variable $x = h/\mu$
  for the parameter set (\ref{modelparameter}), which leads to an
  inflection point at $x = x_c = 0.784$ leading to an extremely flat
  region of the potential. Just before this region, there exists an
  overshooting regime where the universe deviates from the SR
  expansion since the SR parameters are quite large. Moreover,
  curvature perturbations are not frozen at super--horizon scales in
  this regime, as shown in sec.~\ref{srusr}, making a numerical
  treatment necessary.}
\label{potential}
\end{figure}

The inflaton potential for these values of the parameters is shown in
Fig.~\ref{potential}. It features an inflection point\footnote{The
  potential can be expressed in analytical form only in terms of $h$
  or $x$, not in terms of the canonical variable $\chi$. However,
  $dV / dx = d^2 V / dx^2 = 0$ at $x=x_c$ implies
  $d V / d\chi = d^2 V / d \chi^2 = 0$ at $\chi = \chi_c = \chi(x_c)$,
  i.e. the potential of the canonically normalized inflaton also has
  an inflection point. In fact, $V(\chi)$ is qualitatively similar to
  $V(x)$.} at $x = x_c =0.784$. Again following
ref.~\cite{Ezquiaga:2017fvi}, we introduce one more free parameter
$\beta$ in order to study slight deviations from a perfect inflection
point:
\begin{equation}
a \to a(x_c,c ), \ \ b \to (1-\beta) b(x_c,c),
\end{equation}
where $a(x_c,c)$ and $b(x_c,c)$ are the values of the parameters which
lead to $V'(x_c)= V''(x_c)=0$. This is of interest since the inflaton
field can linger near a true inflection point for a very large number
of e--folds. This modification can give a slight slope to the
ultra--flat region. Of course, the shape of the overshooting region,
in particular $V^{\prime \prime}$, will also be slightly modified: the
larger the slope in the ultra--flat region is, the smaller
$V^{\prime \prime}$ will be in the overshooting regime. We will use
$\beta$ in the range $10^{-5}$ to $10^{-4}$.

\subsection{Parameters of the CMB Power Spectrum}
\label{cmbprediction}

The inflaton dynamics in the Einstein frame is given by the
Klein--Gordon equation in curved space:
\begin{equation} \label{inflatondy}
\ddot{\chi} + 3H\dot{\chi} + \frac{dV}{d\chi} = 0 \,.
\end{equation}
Using the relation between the number of e--folds and time,
$dN = H dt$, we can rewrite eq.(\ref{inflatondy}) as
\cite{Ballesteros:2014yva,Ballesteros:2017fsr}
\begin{equation} \label{newreviseddn}
\frac {d^2\chi} {dN^2} + 3 \frac {d\chi} {dN} - \frac{1}{2}
\left( \frac{d\chi}{dN}\right)^3
+ \left[ 3 - \frac{1}{2} \left( \frac {d\chi} {dN} \right)^2 \right]
\frac {V^{\prime}(\chi)} {V(\chi)} = 0\,.
\end{equation}
The two Hubble SR parameters are defined as
\begin{equation} \label{eph}
\epsilon_H = \frac{1}{2} \frac { \dot{\chi}^2} {H^2}
= \frac{1}{2} \left( \frac{d\chi}{dN} \right)^2
\end{equation}
and
\begin{equation} \label{etah}
\eta_{H} = - \frac { \ddot{\chi}} {H \dot{\chi}}
=  \epsilon_H - \frac {1} {2\epsilon_H} \frac {d\epsilon_H}{dN} \,.
\end{equation}
SR inflation requires $\epsilon_H \ll 1$ and $|\eta_H| \ll 1$.

We have seen in sec.~\ref{srusr} that under the SR approximation,
curvature perturbations are basically frozen at super--horizon
scales. The power spectrum is therefore usually computed at horizon
crossing, defined by $k = a H$, and can be analytically given by
\cite{Stewart:1993bc} (see appendix \ref{solums} for details):
\begin{equation} \label{slowrollapp}
\mathcal{P}_{\zeta} \simeq  \frac{H^2}{ 8\pi ^2  \epsilon_H }
\Bigg|_{N=N_{\rm cross}}\,,
\end{equation}
where $N_{\rm cross}$ denotes the number of e--folds at horizon
crossing.\footnote{If $N=0$ defines some initial field configuration,
	only the difference $N - N_{\rm end}$ is well--defined, where
	$N_{\rm end}$ refers to the end of inflation. Successful models have
	to provide at least some $60$ e--folds of inflation, but inflation
	may have lasted much longer.} The scale dependence of $\mathcal{P}_\zeta$
is usually parameterized as a power law, with spectral index $n_s$ given
by
\begin{equation}
n_s - 1  = \frac {d \ln \mathcal{P}_{\zeta} } {d \ln k}
\simeq -4 \epsilon_H + 2 \eta_H\,.
\end{equation}
The deviation from an exact power law is described by the ``running''
of the spectral index, parameterized through the quantity $\alpha$:
\begin{equation}
\alpha = \frac {d n_s} {d\ln k} \simeq \left( -8 \epsilon_H ^2
+ 8 \epsilon_H \eta_H + 2 \frac {d\eta_H} {dN} \right) \,.
\end{equation}
The final CMB observable of phenomenological interest is the tensor to
scalar ratio $r$, i.e. the perturbations in tensor modes (which can be
probed via the polarization of the CMB) normalized to the scalar
perturbations. To leading order in SR parameters,
\begin{equation} \label{tenrsorscalarratio}
r \simeq 16 \epsilon_H\,.
\end{equation}

Now our task is to solve eq.(\ref{newreviseddn}), from which the
parameters of the CMB power spectrum can be computed. We find it more
convenient to calculate the evolution of $x$, rather than the
canonically normalized field $\chi$; this is because we have an
explicit expression for $V(x)$, see eq.(\ref{potential_x}), and thus
also for $V^{\prime}(x)$. By using eq.(\ref{chih}),
eq.(\ref{newreviseddn}) and $x=h/\mu$, we find the differential
equation for $x$ is:
\begin{equation} \label{diffx}
\begin{split}
\mu \left[ \frac {d^2x} {dN^2} g(x)  + \frac {dx} {dN} \frac {dg(x)} {dN}
\right] + 3 \mu \ g(x) \frac {dx} {dN} - & \frac {1} {2} \left( \mu \,
g(x) \frac {dx} {dN} \right)^3 \\
& + \left[ 3 - \frac {1} {2} \left( \mu \, g(x) \frac {dx} {dN} \right)^2
\right] \frac {1} {\mu \, g(x)} \frac {V^{\prime}(x)} {V(x)} = 0\,.
\end{split}
\end{equation}
We have renamed $\frac{d\chi}{dh}(x) \equiv g(x)$ for convenience, see
eq.(\ref{chih}), with $dg / dN = (dx/dN) dg/dx$. Eq.(\ref{diffx}) is
too complicated to solve analytically. For a numerical solution we
have to choose initial values for $x$ and $dx/dN$ at some $N =
N_0$. The initial value of $x$ should evidently be above the field
values where the CMB scales cross the horizon, so that our solution
covers the entire range of scales probed by the CMB and other
cosmological observations. On the other hand, it would be wasteful to
choose $x(N_0)$ to be much larger than the field values probed by the
CMB, since this earlier evolution leaves no observable traces
anyway. In practice we have used $x(0) = 8.5$. At these large field
values the potential is very flat; if the initial kinetic energy of
the field is not very large, the field evolution will therefore
quickly approach the SR solution.\footnote{In other words, SR
	is a strong attractor solution of the equation of motion when going
	forward in time. This also implies that one practically cannot solve
	this equation going backward in time: for almost all initial
	conditions the solution for $x$ will then quickly blow up.} The
initial choice of $dx/dN$ is therefore largely irrelevant; we chose
$dx/dN = -0.21$, which corresponds to assuming the SR solution already
at $N = N_0 = 0$.

With these initial conditions, eq.(\ref{diffx}) can be then
solved numerically. Once $x(N)$ is known, the evolution of the
canonical field $\chi$ can be obtained by integrating
eq.(\ref{chih}):
\begin{equation}
\chi(N) = \int_{0}^{N} \mu \, g(x(N^{\prime}))\, \frac{dx}{dN^{\prime}}\,
dN^{\prime} + \chi(0)\,.
\end{equation}
The constant of integration $\chi(0)$ can be fixed by using the fact
that $g(x) \to 1$ for $x \to 0$; the natural choice is thus
$\chi \simeq h$ for $\xi h \ll 1$, which corresponds to
$\chi(0) = 6.94$.

\begin{figure}[h!]
\centering
\includegraphics[width=0.6\paperwidth, keepaspectratio]{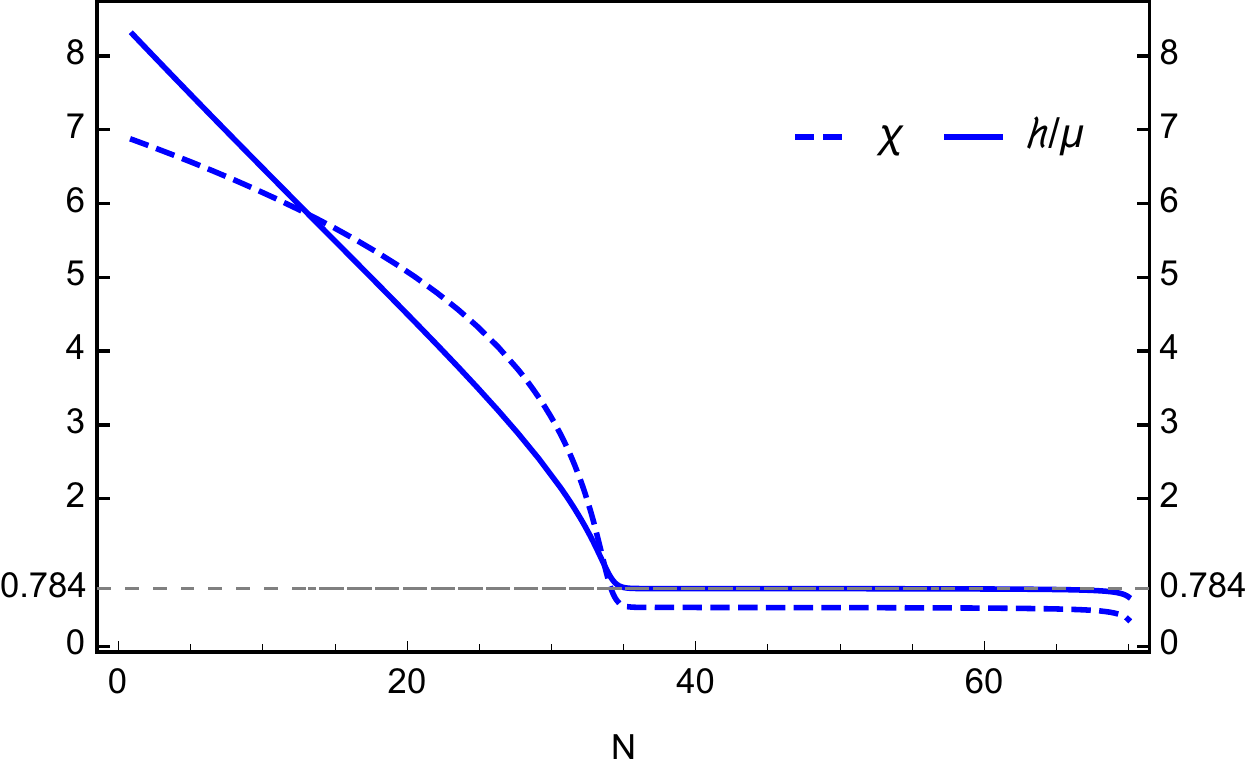}
\caption{Evolution of the Higgs field with $N$. The universe expands
	by more than 30 e--folds while the inflaton field traverses the flat
	region of the potential around the pseudo--critical point
	$x = x_c = 0.784$; this corresponds to the USR phase. Between SR and
	USR, there is an intermediate overshooting stage where the
	canonically normalized inflaton field $\chi$ varies rather quickly
	with $N$.}
\label{higgs}
\end{figure}

Fig.~\ref{higgs} shows the evolution of $x$ as well as the canonically
normalized field $\chi$ with $N$ for our standard set of parameters
(\ref{modelparameter}) with $\beta = 10^{-5}$. We see that the field
at first gradually decreases with increasing $N$; this is the usual SR
phase, for large field values. The evolution of $\chi$ becomes quite
fast at $N \simeq 30$, signaling a break--down of SR. However, from
$N \simeq 36$ both $x$ and $\chi$ remain nearly constant for more than
$30$ e--folds; this is when the inflaton traverses the very flat part
of the potential around the pseudo--critical point. Evidently the
behavior of the field differs qualitatively from that in the SR phase,
justifying the use of the expression ``ultra--slow roll'' for much of
this epoch. Inflation ends when the inflaton leaves this region.

Once the dynamics of the inflaton field is known, the parameters of
the CMB power spectrum can be computed. Using our standard parameter
set (\ref{modelparameter}) and $\beta =10^{-5}$ we find that the CMB
``pivot scale'' $k=0.05 \ \mathrm{Mpc^{-1}}$ crosses out of the
horizon at $N_{\rm end} - N_{\rm CMB} \approx 68$. The numerical values
of the CMB parameters at this scale are
\begin{equation} \label{cmbpre}
\mathcal{P}_\zeta = 2.09 \times 10^{-9};\  n_s = 0.951; \  \alpha =- 0.0018;\
r =0.043\,,
\end{equation}
which is consistent with the Planck 2018 results \cite{Akrami:2018odb}
at the $2\sigma $ level. The large value of
$N_{\rm end} - N_{\rm CMB}$ is to a large extent due to the USR
phase. This number of e--folds can be reduced by increasing $\beta$,
which increases the slope of the potential near the pseudo--critical
point. For example, using $\beta = 10^{-4}$, we find the same
predictions as given by eq.(\ref{cmbpre}) at
$N_{\rm end} - N_{\rm CMB} \approx 63$.

\subsection{Slow-roll Violation}
\label{slowrollvio}

For our standard set of parameters, CMB scales first crossed out of
the horizon during a SR phase, i.e. the SR approximation works
very well for the predictions collected in eq.(\ref{cmbpre}).
However, Fig.~\ref{higgs} also shows that the canonically normalized
inflaton field $\chi$ moves rather fast for $N \simeq 33$. In this
subsection we show that the SR conditions are indeed violated in this
``overshooting'' region.

\begin{figure}[htbp]
\centering
\subfloat[]{\includegraphics[width=75mm]{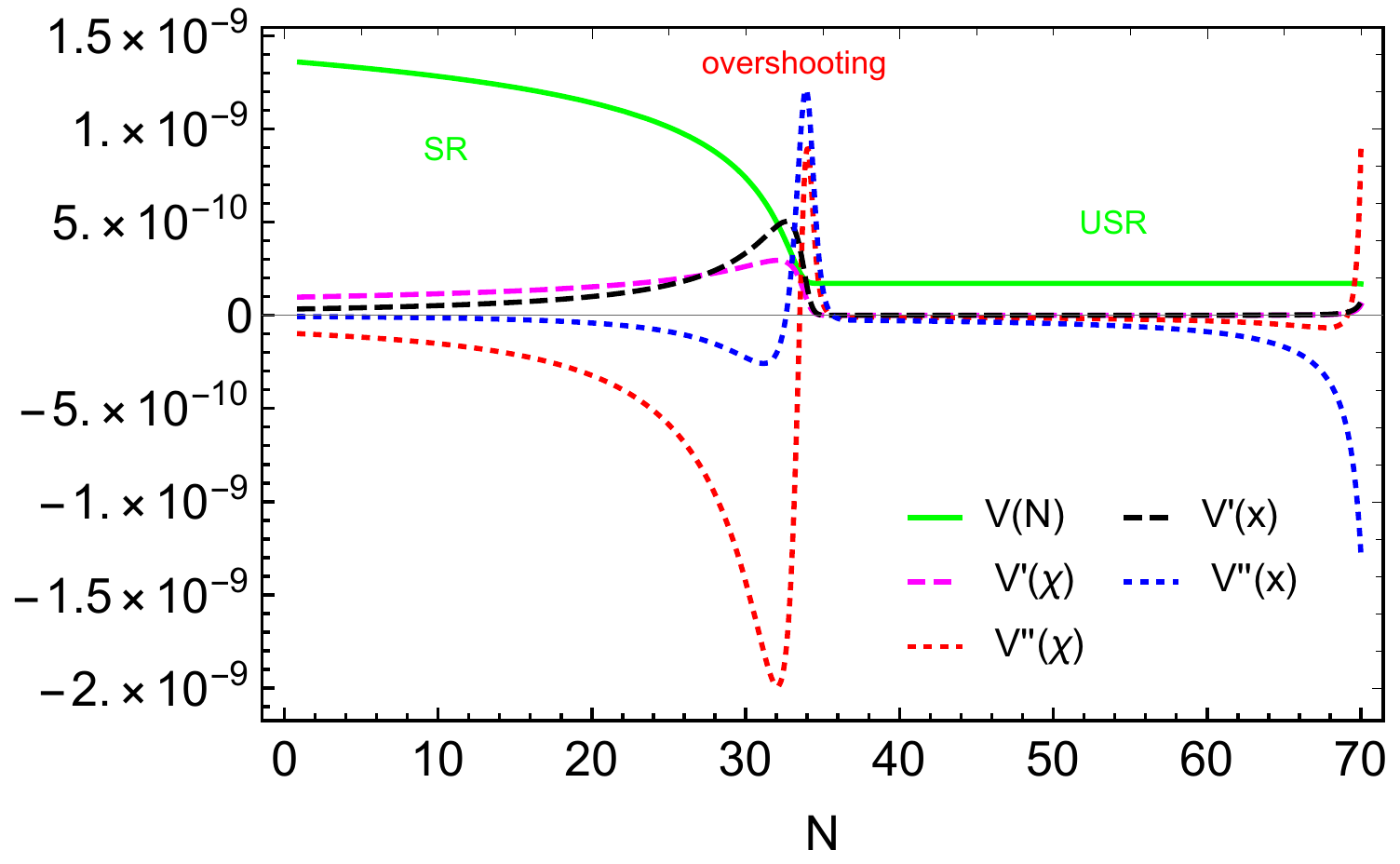}
	\label{potentialvsn}	}
\subfloat[]{\includegraphics[width=70mm]{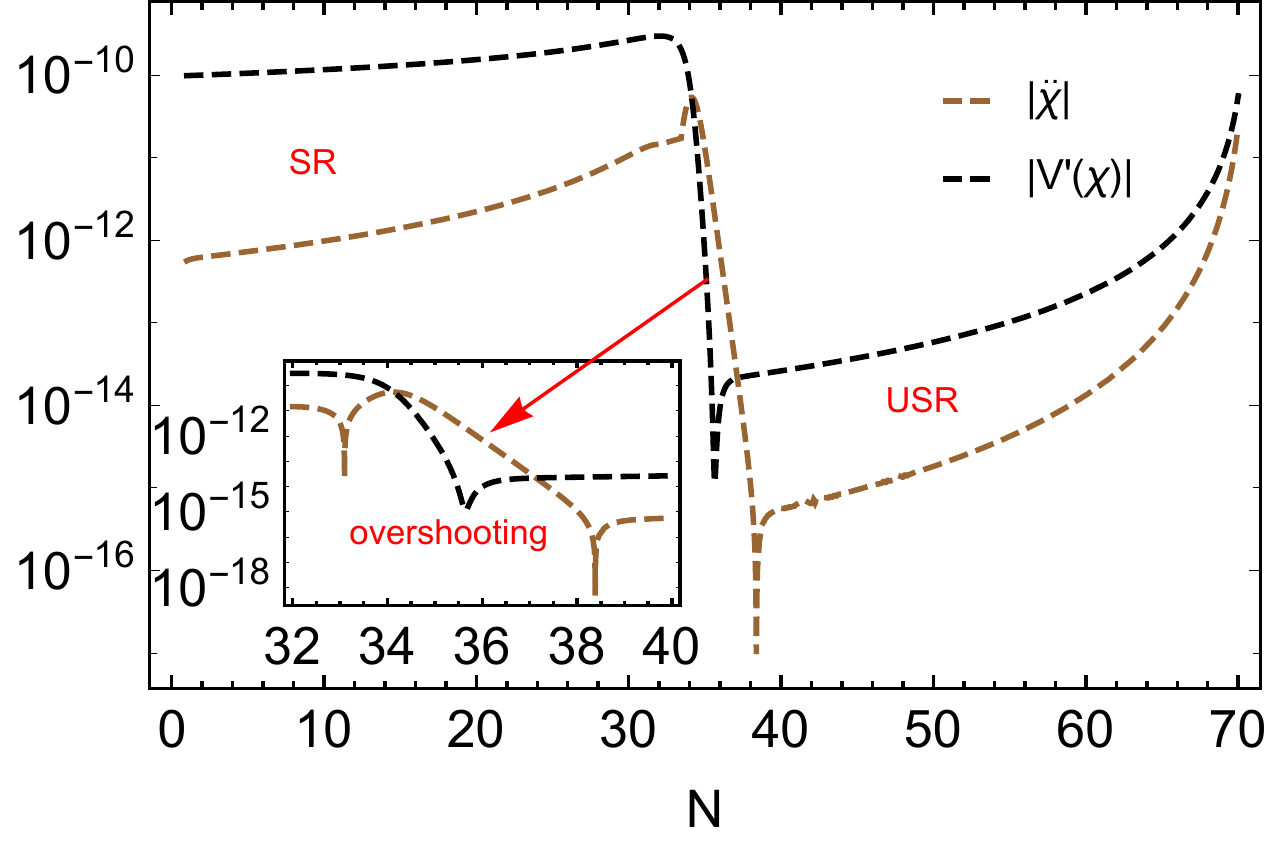}
	\label{overshooting}	}
\caption{The left frame shows the evolution of potential and its
	derivatives with $N$. There is a first SR phase at $N < 30$ where
	all derivatives are small, and a USR phase at $N > 38$ where the
	derivatives are indistinguishable from zero on the shown scale.  In
	between there is an overshooting stage where the curvature of the
	potential is rather large and varies rapidly; in this regime the SR
	approximation breaks down and entropic perturbations are
	excited. This is further illustrated in the right frame which
	compares the second time derivative $\ddot\chi$ of the canonically
	normalized inflaton field with the slope of the potential; in the
	overshooting region, $N \sim 35$, the former considerably exceeds
	the latter in magnitude.}
\label{fieldbackgroun}
\end{figure}

The dependence of the potential and its first and second derivatives,
both with respect to $x$ and with respect to $\chi$, are plotted as
function of $N$ in Fig.~\ref{potentialvsn}. The first derivatives
remain positive and fairly small throughout. The second derivatives
are initially small and negative, but increase in size as the inflaton
field approaches the overshooting region, where the second derivative
changes very rapidly from large negative to large positive values; in
the region around the pseudo--critical point the second derivatives
are again very small.

When discussing non--adiabatic pressure perturbations in Sec.~2, we had
assumed that the second time derivative of the inflaton field is much
larger in magnitude than the slope of the potential. Fig.~\ref{overshooting}
shows that this is indeed the case for some range of $N$ around $35$.
In this case eq.(\ref{inflatondy}) becomes
\begin{equation}
\ddot{\chi} + 3H \dot{\chi} \approx 0,
\end{equation}
which implies 
\begin{equation} \label{etah3}
\eta_H = - \frac{\ddot{\chi}}{H\dot{\chi}} \approx 3.
\end{equation}
In SR, both $\epsilon_H$ and $|\eta_H|$ should be (much) smaller
than $1$; the result (\ref{etah3}) clearly violates this.

\begin{figure}[h]
\centering
\includegraphics[width=.6\paperwidth, keepaspectratio]{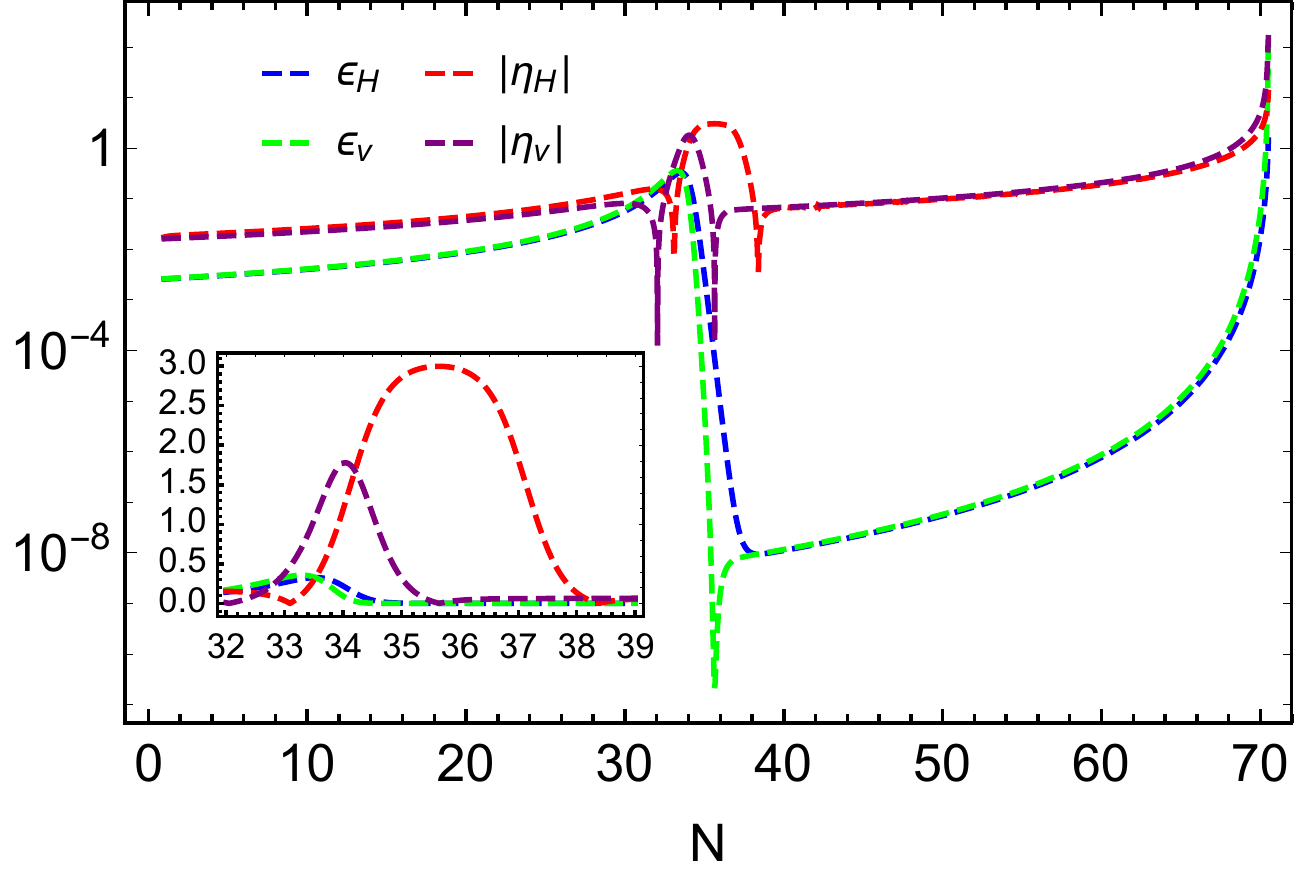}
\caption {The evolution of the SR parameters with $N$. The parameters
  defined via the potential almost coincide with those defined via the
  Hubble parameter in the epochs where the SR approximation holds, but
  they differ markedly in the overshooting region where $\eta_H > 1$.}
\label{hubbleslowroll} 
\end{figure}

This is further illustrated in Fig.~\ref{hubbleslowroll}, which shows
the evolution of the SR parameters with $N$ for the same set of
parameters. We show both the ``Hubble'' SR parameters defined
in eqs.(\ref{eph}) and (\ref{etah}) and their ``potential'' analogues,
defined via
\begin{equation}
\epsilon_V = \frac{1}{2} \left( \frac {V'}{V} \right)^2\,, \ \
\eta_V = \frac {V^{\prime\prime}} {V}\,.
\end{equation}
SR implies that $\epsilon_H \simeq \epsilon_V$ and
$\eta_H \simeq \eta_V$; we see that in our case this holds for
$N < 30$ as well as\footnote{In our example these relations even hold
  at $N \simeq 70$ where SR no longer holds since inflation ends.} for
$N > 38$. $\epsilon_H$ and $\epsilon_V$ always remain significantly
smaller than $1$, but vary rapidly, and differ markedly, in the
overshooting region.\footnote{We saw in Fig.~\ref{higgs} that the
  inflaton already moves very slowly at $N=36$. However, since all the
  SR parameters, in particular $\eta_H$, become small only for
  $N \geq 38$, we denote only this epoch as USR epoch. Defining USR
  via the condition $\eta_H \simeq 3$, as seems to be done in part of
  the literature, does not seem very useful to us, since at the
  beginning of the epoch where this condition is satisfied the
  inflaton field still mover rather quickly; conversely, for much of
  the time where the inflaton moves extremely slowly, $\eta_H \ll
  1$. We instead use $\eta_H \simeq 3$ to define the overshooting
  region.}

We have shown in sec. \ref{srusr} that entropic perturbation can be
excited if the SR conditions are violated, in which case the curvature
perturbation are no longer conserved at super horizon scale. Hence the
usual estimate (\ref{slowrollapp}) of the power spectrum can no longer
be justified for modes that first crossed out of the horizon near the
overshooting region. In the next section we instead use the
Mukhanov--Sasaki (MS) formalism to compute the power spectrum
numerically.

\subsection{Mukhanov--Sasaki Formalism}

Our numerical treatment of the evolution of the curvature
perturbations is based on the MS equation; a quick
derivation of this equation and its analytical solution in the quasi--de
Sitter limit is reviewed in appendix \ref{solums}. It is usually written
in terms of the Mukhanov variable
\begin{equation} \label{mkvariable}
v_k \equiv -z\zeta_k\,,
\end{equation}
where $z$ is defined by
\begin{equation}  \label{zdef}
z^2 \equiv \left( \frac {d\chi}{dt} \right)^2 \frac{a^2 }{H^2}
= 2a^2 \epsilon_H\, .
\end{equation}
In these variables, the MS equation reads:
\begin{equation} \label{MukhanovSasaki}
\frac {d^2v_k} {d\tau^2} + \left( k^2-\frac{1}{z} \frac{d^2z}{d\tau^2}
\right) v_k = 0,
\end{equation}
where $\tau$ denotes the conformal time, i.e. $d\tau = \frac {dt}{a}$.

Rewriting eq.(\ref{MukhanovSasaki}) using the number of e--folds $N$ instead
of the conformal time gives \cite{Ballesteros:2017fsr}
\begin{equation} \label{remuk}
\begin{split}
\frac {d^2v_k} {dN^2} &+ ( 1 -\epsilon_H) \frac {dv_k} {dN} \\
&+ \left[ \frac {k^2} {a^2H^2} + ( 1 + \epsilon_H - \eta_H )
( \eta_H - 2 ) - \frac {d(\epsilon_H-\eta_H)} {dN} \right] v_k = 0 \,.
\end{split}
\end{equation}
Under SR conditions the curvature perturbation $\zeta_k$ is frozen at
super--horizon scales, hence the power spectrum of $\zeta_k$ is usually
computed at horizon crossing. However, we have seen in the previous
subsection that the SR approximation fails in the overshooting regime.
In order to account for the evolution of the curvature perturbation
also at super--horizon scales the power spectrum should be computed at
the end of inflation:
\begin{equation} \label{msnend}
\mathcal{P}_\zeta(k) = \frac {k^3} {2\pi^2} \Big| \zeta_k\Big|^2_{N=N_{\rm end}}
= \frac {k^3} {2\pi^2} \Big| \frac {v_k} {z} \Big|^2_{N=N_{\rm end}}\,.
\end{equation}
This can usually only be done numerically. Recall also that
super--horizon perturbations are frozen after inflation, as we showed
at the very end of Sec.~2.

In order to solve eq.(\ref{remuk}), initial conditions are needed. We
follow the usual procedure, which assumes the Bunch--Davies vacuum at
very early times \cite{Bunch:1978yq}:
\begin{equation} \label{initialconds}
\lim\limits_{\tau \to -\infty} v_k = \frac {{\rm e}^{-ik\tau}} {\sqrt{2k}}\,.
\end{equation}
Since $v_k$ is a complex quantity,\footnote{The perturbation $\zeta$
	introduced in eq.(\ref{gaugeinvariant}) is a real quantity in
	configuration space, but its Fourier coefficients $\zeta_k$ are in
	general complex.} in practice it is more convenient to solve for its
real and imaginary parts separately. To this end one can rewrite the
initial condition eq.(\ref{initialconds}) as
\cite{Ballesteros:2017fsr}:
\begin{equation} \label{reini}
\mathrm{Re}(v_k) \Big|_{N=N_i} = \frac {1} {\sqrt{2k}};  \
\mathrm{Im}(v_k) \Big|_{N=N_i} = 0;
\end{equation}
\begin{equation} \label{imini}
\mathrm{Re} \left( \frac {dv_k} {dN} \right) \Big|_{N=N_i} =0; \
\mathrm{Im} \left( \frac {dv_k} {dN} \right) \Big|_{N=N_i} =
- \frac {\sqrt{k}} {\sqrt{2}a(N_i) H(N_i)}\,.
\end{equation}
Here $N_i$ is the ``initial'' point where we start the numerical
integration of the MS equation. In principle the Bunch--Davis initial
conditions (\ref{initialconds}) should be imposed at
$\tau \to -\infty$, which also corresponds to $N \to - \infty$ if the
CMB pivot scale crossed the horizon at $N \simeq 0$, as we assumed in
the last three figures.  Physically this does not make much sense,
since we don't know how many e--folds of inflation happened before
that time. Moreover, the ansatz (\ref{initialconds}) remains a very
good approximation of the exact solution of the MS equation as long as
the perturbation is well within the horizon, i.e. for
$k^2 \gg a^2 H^2$.  Let the mode $k$ cross the horizon at
$N = N_{k,{\rm cross}}$. In practice it is then usually sufficient to
use $\Delta N \equiv N_{k,{\rm cross}}- N_i= 2 \sim 3$. We have
checked that reducing $N_i$, which costs a lot of CPU time since the
term $k^2/(a^2H^2)$ in eq.(\ref{remuk}) grows
$\propto {\rm e}^{2 \Delta N}$ requiring correspondingly reduced step
sizes to attain numerical convergence, does not change the final
result appreciably. However, we will see below that choosing too small
a value for $\Delta N$ can lead to inaccuracies.

\subsection{Power Spectrum for Critical Higgs Inflation}

We now apply the MS formalism to CHI. In order to compute the power
spectrum we have to integrate eq.(\ref{remuk}) with the initial
conditions eq.(\ref{initialconds}) till the end of inflation, and then
plug the solution into eq.(\ref{msnend}).

\begin{figure}[htbp]
\centering
\includegraphics[width=.6\paperwidth, keepaspectratio]{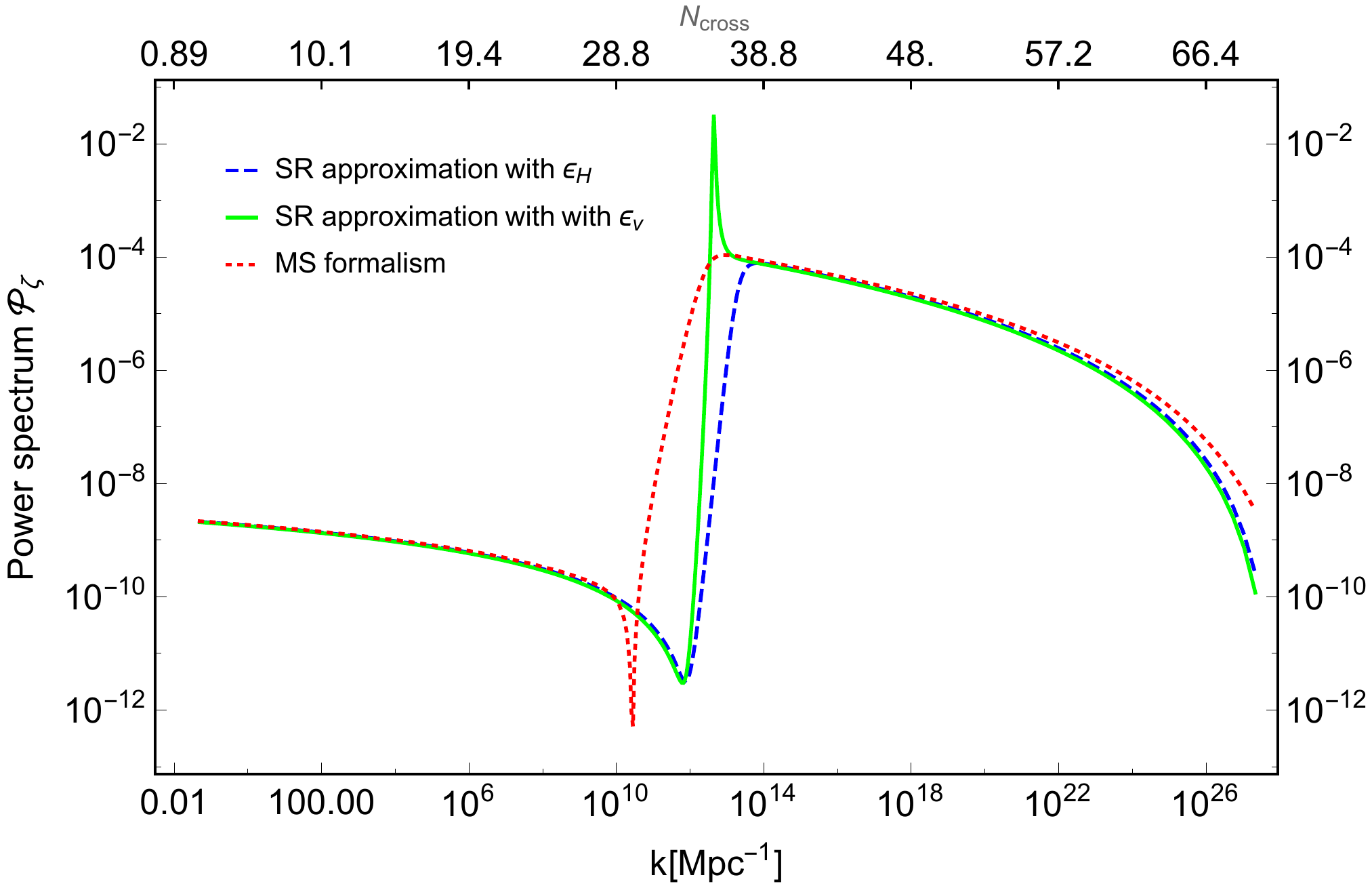}
\caption {Comparison of the power spectrum computed using the SR
  approximation (the blue dashed curve corresponds to
  eq.(\ref{slowrollapp}) while the green line represents results with
  the replacement $\epsilon_H \to \epsilon_V$) and the MS formalism
  (red dotted).}
\label{ms}
\end{figure} 

The result for our standard parameter set with $\beta = 10^{-5}$ is
shown in Fig.~\ref{ms}. We see that the SR approximation fails badly
for modes crossing the horizon in the vicinity of the overshooting
region. In particular, the SR approximation gets both the location and
the depth of the dip in the power spectrum wrong by more than one
order of magnitude. The approximation (\ref{slowrollapp}) using
$\epsilon_H$ underestimates the maximum of $\mathcal{P}_\zeta$ by only
a factor of about 1.4, but gets the location
$k_{\rm max}$ of the true maximum off by an order of magnitude, and
underestimates the power at $k_{\rm max}$ by about five orders of
magnitude. Using the approximation (\ref{slowrollapp}) but replacing
$\epsilon_H$ by $\epsilon_V$, as is done in much of the older
literature on inflation, actually gets $k_{\rm max}$ approximately
right, but overestimates the power at this scale by more than two
orders of magnitude. In contrast, the SR approximation works well both
for the large scales probed by the CMB and for the much smaller scales
that cross out of the horizon in the USR regime after the end of the
overshooting epoch. The power at these small scales exceeds that at
CMB scales by roughly five orders of magnitude due to the overshooting behavior \footnote{Regarding the jump  of the power spectrum, we thank the anonymous referee for bringing refs. \cite{Starobinsky:1992ts,Ivanov:1994pa,Hazra:2014jka,Hazra:2014goa,Hazra:2017joc} to our attention. These papers consider some discontinuous step  in the inflaton potential, which can give rise to interesting wiggles in the  power spectrum. Depending on regime of the discontinuity (motivated by \cite{Starobinsky:1992ts}), the resulting primordial power spectrum  can lead to significant production of primordial black hole \cite{Ivanov:1994pa}, and can even offer better fit for the Planck data with the so-called Wiggly Whipped Inflation model \cite{Hazra:2014jka,Hazra:2014goa,Hazra:2017joc}, where an overshooting phase can also appear.}.

\begin{figure}[h!]
\centering
\subfloat[\small $k=2.5\times10^{10} \ \mathrm{Mpc^{-1}} $, $N_{\rm cross}=29.7$]{
\includegraphics[width=70mm]{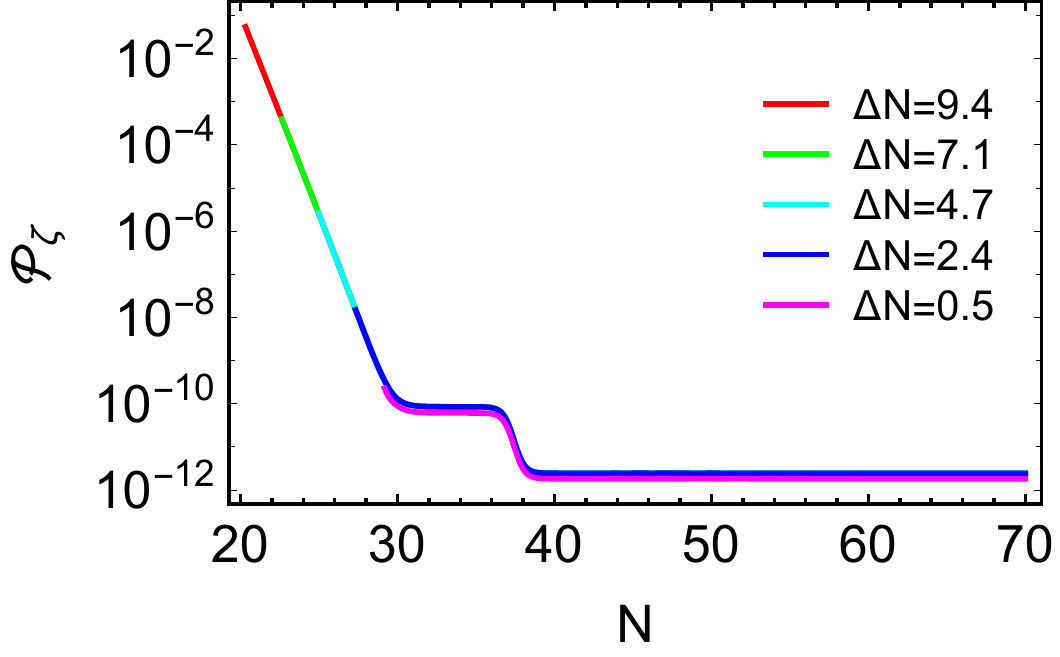}
\label{plotk10} }
\subfloat[\small $k=10^{11} \ \mathrm{Mpc^{-1}} $ , $N_{\rm cross}=31.2$]{
\includegraphics[width=70mm]{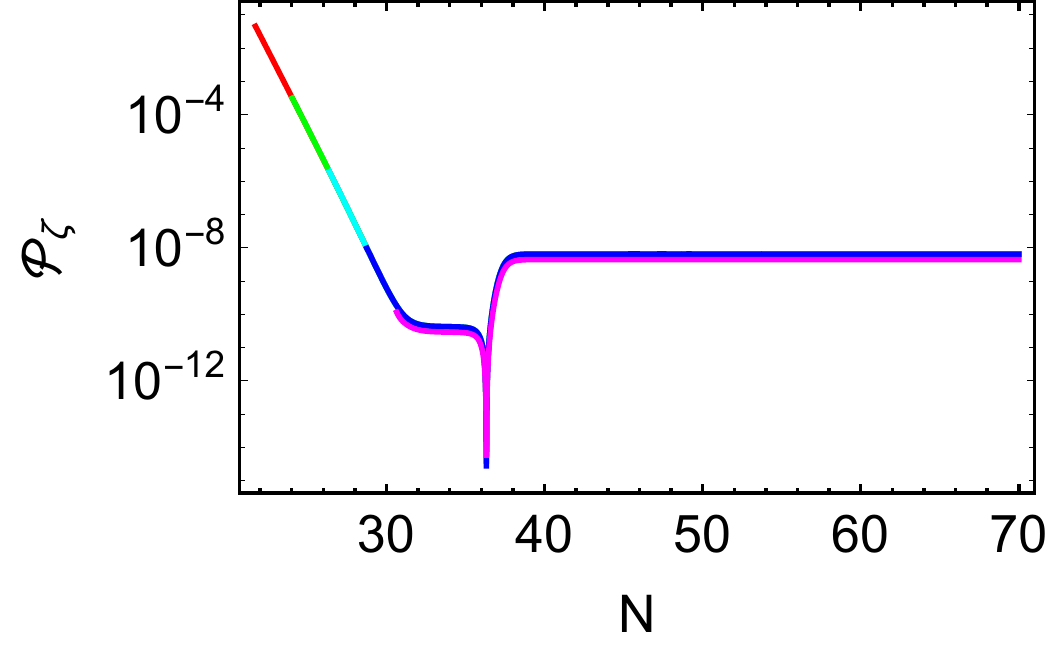}
\label{plotk11}}
\hspace{0mm}
\subfloat[\small $k=10^{13} \ \mathrm{Mpc^{-1}} $ , $N_{\rm cross}=36.5$]{
\includegraphics[width=70mm]{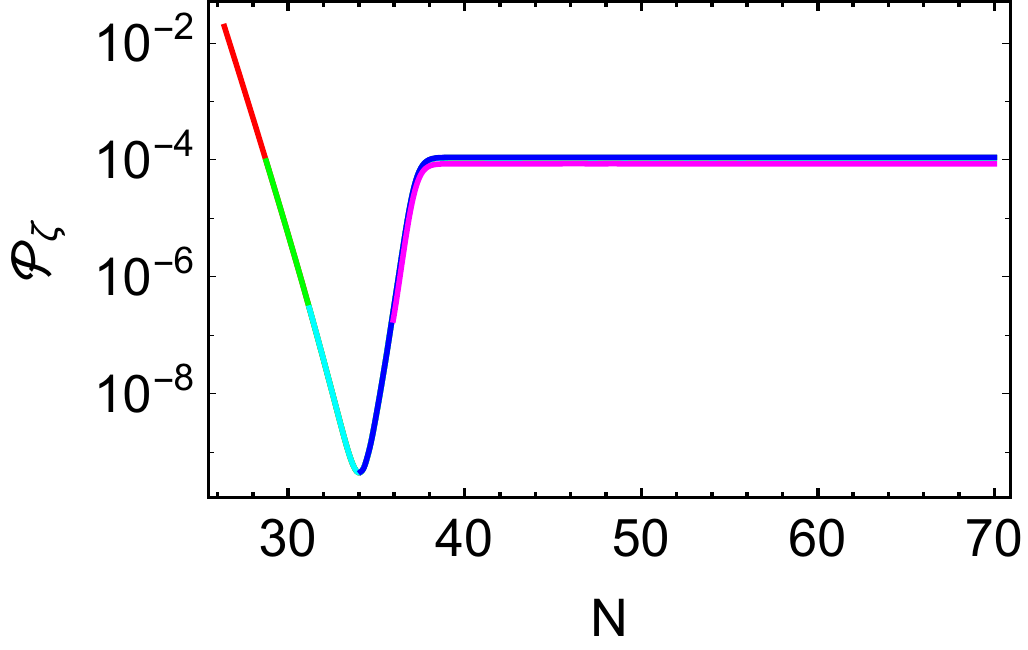}
\label{plotk13} }
\subfloat[\small $ k= 2\times 10^{14} \ \mathrm{Mpc^{-1}} $ , $N_{\rm cross}=39.5$]{
\includegraphics[width=70mm]{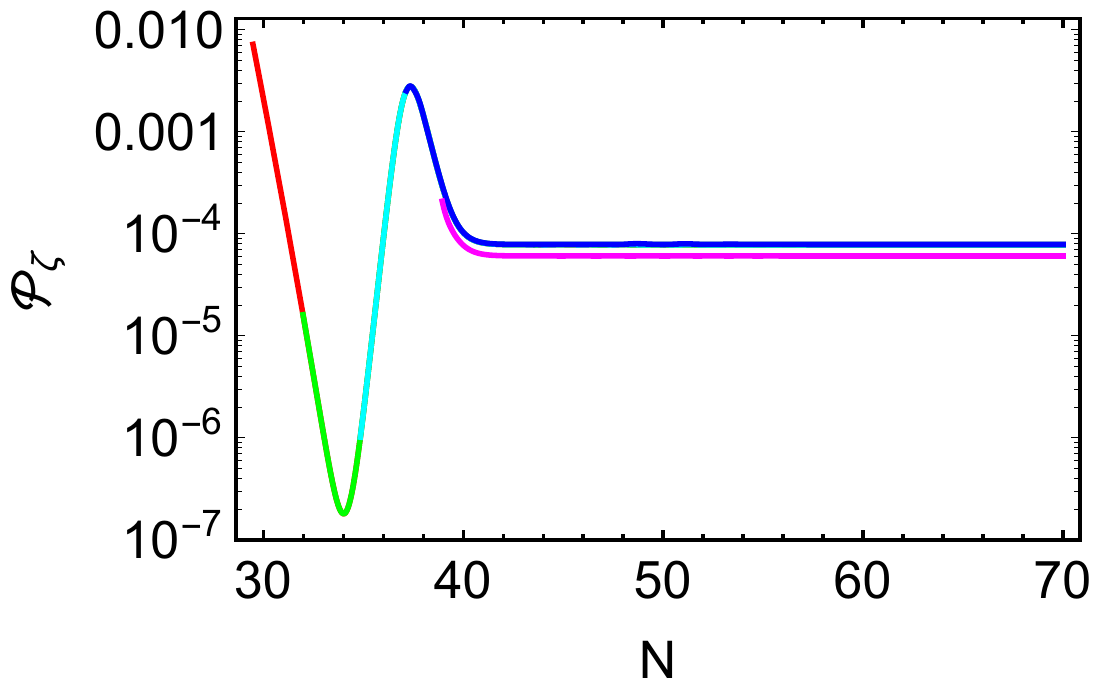}
\label{plotk14} }
\caption {Evolution of the power spectrum for four different modes
  which cross the horizon at $N_{\rm cross}$ near the overshooting
  region. The different colors refer to different initial values
  $N_i = N_{\rm cross} - \Delta N$ when integrating eq.(\ref{remuk}).
  Evidently it is sufficient to use $\Delta N = 2 \sim 3$ in practice,
  since the curves for even smaller $N_i$ merge with each other, and
  yield same results. However, using too small $\Delta N$ leads to
  inaccurate results, for example the one we have showed with
  $\Delta N =0.5$.}
\label{differentk}
\end{figure}

In order to better understand the red curve in Fig.~\ref{ms}, in
Fig.~\ref{differentk} we show the evolution of $\mathcal{P}_\zeta$
with $N$ for four representative values of $k$. These results have
been obtained by numerically solving the MS equations; the different
curves refer to different values of $N_i = N_{\rm cross} - \Delta N$
where the initial conditions (\ref{initialconds}) have been imposed.

While the results of Fig.~\ref{differentk} have been obtained from
eq.(\ref{remuk}), the qualitative behavior is more easily understood by
combining eqs.(\ref{MukhanovSasaki}) and (\ref{mkvariable}), which
yields the equivalent differential equation
\begin{equation} \label{superevo}
\frac{d^2\zeta_k}{dN^2} + ( 3 + \epsilon_H - 2\eta_H) \frac {d\zeta_k} {d N}
+ \frac {k^2} {a^2 H^2} \zeta_k = 0\,.
\end{equation} 
This equation again has to be satisfied by both the real and imaginary
parts of $\zeta_k$.

For sub--horizon modes, where $k^2 \gg a^2 H^2$, the last term in
eq.(\ref{superevo}) dominates; this by itself leads to an oscillatory
behavior of $\zeta_k$, with amplitude increasing
$\propto {\rm e}^{N/2}$ and with exponentially decreasing oscillation
frequency. For SR conditions, $\epsilon_H, \, |\eta_H| \ll 1$,
the second term in eq.(\ref{superevo}) is a damping term, which
reduces the amplitude of the oscillations $\propto {\rm
	e}^{-3N/2}$. Altogether this yields
$\mathcal{P}_\zeta \propto {\rm e}^{-2N}$, which explains the initial
steep decline in all four cases depicted in Fig.~\ref{differentk}.

Of course, the term $\propto k^2$ in eq.(\ref{superevo}) decreases
$\propto {\rm e}^{-2N}$, due to the exponential growth of $a(N)$; by
definition the coefficient multiplying $\zeta_k$ in this term equals
$1$ at $N = N_{\rm cross}$. Moreover, the SR conditions are badly
violated in the overshooting region. The evolution of the power
depends on where $N_{\rm cross}$ lies relative to the overshooting
region.  To see this, let us discuss the four cases depicted in
Fig.~\ref{differentk} one by one.

Fig.~\ref{plotk10}: Here we chose
$k = 2.5\times 10^{10} \ \mathrm{Mpc^{-1}}$, so that horizon crossing
takes place at $N=29.7$, where the SR conditions still hold. As shown
in Fig.~\ref{plotk10}, the power spectrum for this mode first
approaches a constant after horizon crossing. Here the last term in
eq.(\ref{superevo}) is negligible. As long as the coefficient of the
second term is close to $+3$, the absolute value of the first
derivative of $\zeta_k$ keeps decreasing exponentially with increasing
$N$; this corresponds to an overdamped oscillator. The solution for
this range of $N$ can thus be written as
$\zeta_k(N) = C_1 + C_2 e^{-3 (N - N_{\rm cross} )}$, where $C_1, C_2$
are two constants determined by the initial conditions\footnote{The
	value of $C_2$ is roughly of order $\mathcal{O}(10^{-5})$ according
	to our finding in eq.(\ref{zetadot2}), while $C_1$ depends on
	$k$.}. Let $N_{\rm SR}$ denote the number of e--folds which $\zeta_k$
undergoes in the SR regime after horizon crossing, but before
overshooting; then this epoch suppresses the first derivative of
$\zeta_k$ by a factor ${\rm e}^{-3 N_{\rm SR}}$. Since the derivative of
$\zeta_k$ is small, $\zeta_k$ itself is basically constant.

This solution is no longer valid in the overshooting region, where
$\eta_H \approx 3$ while $\epsilon_H$ remains rather small, so that
$(3+\epsilon_H-2\eta_H) \approx -3$, i.e. the second term in
eq.(\ref{superevo}) changes sign relative to the SR epoch. This means
that now the first derivative to $\zeta_k$ begins to {\em grow}
exponentially in magnitude, however without changing sign. At the end
of this epoch one thus has
$\zeta_k(N) = C_3 + C_4 {\rm e}^{+3N_{\rm OS}}$, where $N_{\rm OS}$ is
the total ``length'' of the overshooting epoch, i.e. the number of
e--folds during which $\eta_H \approx 3$.\footnote{The exponential
	growth of $d \zeta_k / dN$ agrees with our earlier discussion of
	eq.(\ref{curvature_evolution}).} By matching the two solutions for
$\zeta_k$ at the point where the overshooting epoch begins, one finds
$C_1 =C_3$ and $C_4 = - C_2 {\rm e}^{-3N_{\rm SR}}$. So after
overshooting ends, the value of $\zeta_k$ is approximately given by
$\zeta_k = C_1 - C_2 e^{3\left( N_{\rm OS} - N_{\rm SR} \right)
}$. Since afterwards the SR conditions hold again, $\zeta_k$ remains
approximately constant, i.e. this result still holds at the end of
inflation.\footnote{Actually after the overshooting dynamics ends, the
	inflaton enters the USR phase, where the matter perturbation is even
	more adiabatic compared to that in a SR.}  

\begin{figure}[h!]
\centering
\subfloat[Evolution of $A$.]{\includegraphics[width=70mm]{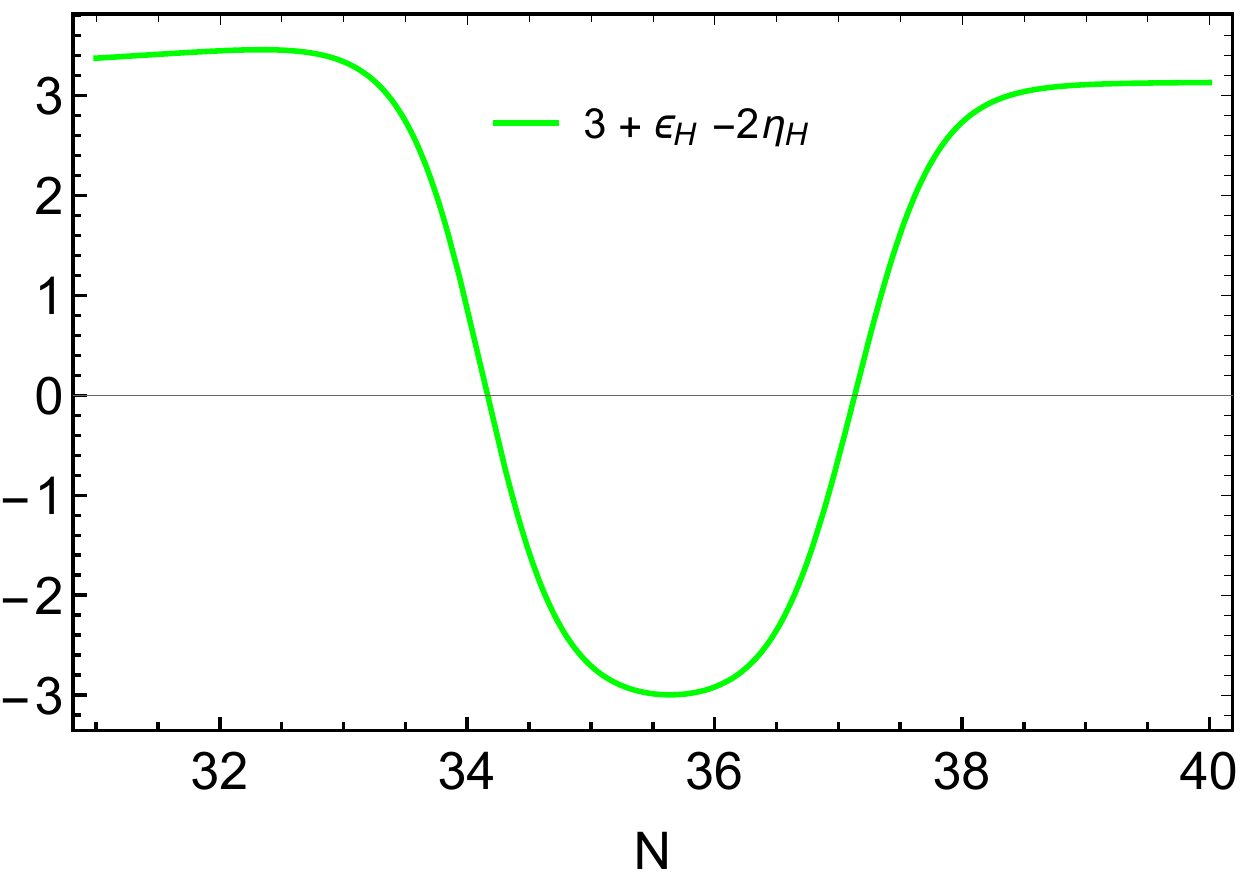} \label{plota} }
\subfloat[Evolution of $B$.]{\includegraphics[width=72mm]{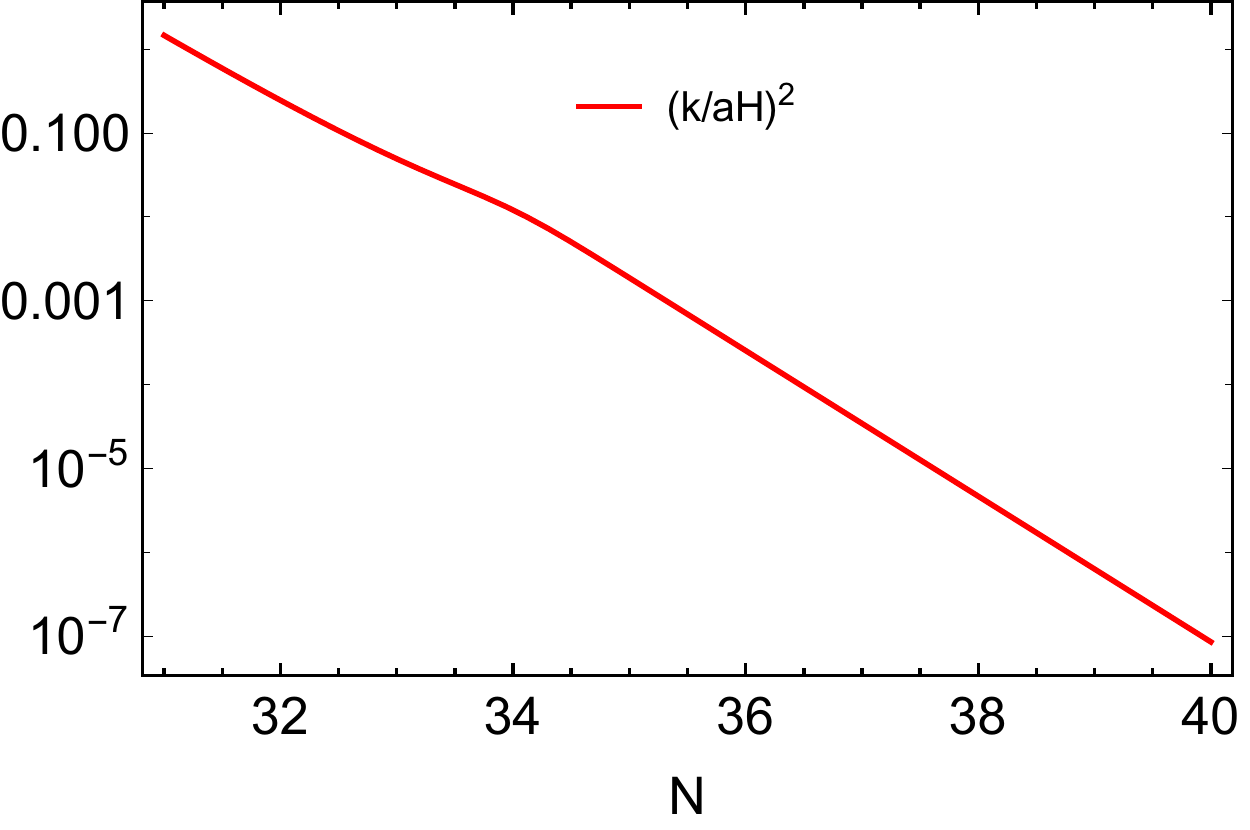} \label{plotb} }
\caption {Evolution of the two coefficients in eq.(\ref{superevo});
  the left frame is independent of $k$, while the result of the right
  holds for $k= 10^{11} \ \mathrm{Mpc^{-1}}$ and scales
  $\propto k^2$.}
\label{ab}
\end{figure}

The overshooting region will therefore only have significant impact on
the final power for modes that crossed out of the horizon not much
more than $N_{\rm OS}$ e--folds before its onset. From the left frame
of Fig.~\ref{ab} we read off that for our numerical example
overshooting starts at $N \approx 34$, with $N_{\rm OS} \approx
3.5$. For the case considered in Fig.~\ref{plotk10} $N_{\rm OS}$ and
$N_{\rm SR}$ are comparable. For much smaller co--moving wave number
$k$, $N_{\rm SR} \gg N_{\rm OS}$, so that the effect of the
overshooting region on the final power is not significant. This explains
why the standard SR approximation works for $k < 10^{10} \ \mathrm{Mpc^{-1}}$
in Fig.~\ref{ms}.

\begin{figure}[h!] 
\centering
\includegraphics[width=.6\paperwidth, keepaspectratio]{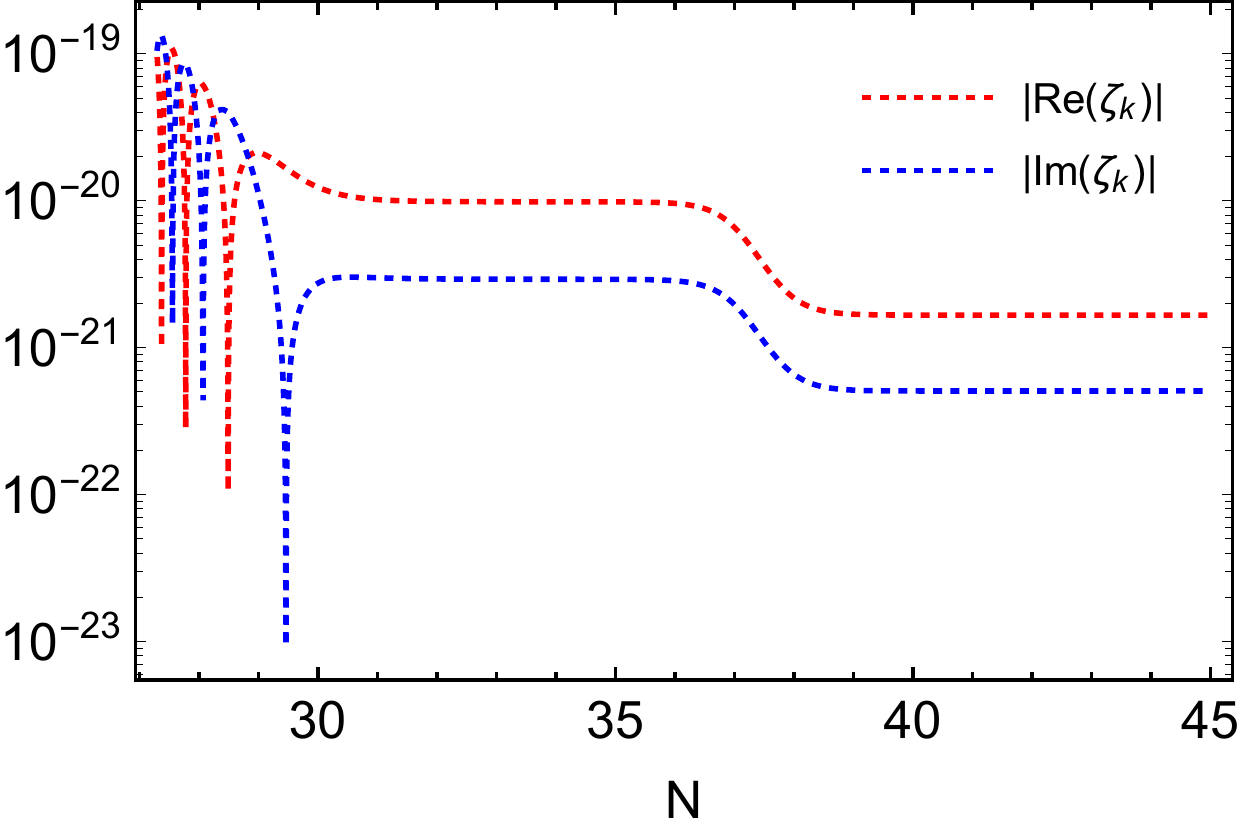}
\caption {Evolution of curvature perturbations for the mode with
  co--moving $k= 2.5\times 10^{10} \ \mathrm{Mpc^{-1}} $. This figure
  shows the epoch from a few e--folds before horizon crossing to a few
  e--folds after the end of the overshooting region.}
\label{xi_k10}
\end{figure} 

The detailed evolution of the real and imaginary parts of $\zeta_k$ is
shown in Fig.~\ref{xi_k10}. Note that the overshooting region has
significant impact on $\zeta_k$ itself (as opposed to its derivative)
only beginning at $N \approx 37$, where the exponential growth of the
modulus of the derivative compensated its exponential suppression
between horizon crossing and the onset of the overshooting
epoch. Since overshooting already ends at $N \approx 38$, its total
effect is still moderate for this value of $k$. Notice, however, that
the second flat region lies well below the first one, which
corresponds to the prediction of the usual analytical SR
estimates. This is because in the overdamped oscillator phase just
after horizon crossing, the first derivative of $\zeta_k$ always has the
opposite sign as $\zeta_k$ itself, for both the real and imaginary part.
The exponential decrease of the modulus of the derivatives will therefore
decrease $|\zeta_k|$, and thus $\mathcal{P}_\zeta$. We will come back to
this point shortly.

Fig.~\ref{plotk11}: for co--moving $k=10^{11} \ \mathrm{Mpc^{-1}}$,
$\mathcal{P}_\zeta$ nearly vanishes for a value of $N$ during the
overshooting epoch.\footnote{To the best of our knowledge, a similar
	behavior as shown in Fig.~\ref{plotk11} was first explored in
	\cite{Leach:2000yw} and recently was mentioned in
	\cite{Cicoli:2018asa, Ezquiaga:2018gbw}.} Now the nominal horizon
crossing at $N = 31.2$ occurs just before the onset of the
overshooting epoch, which means we cannot always assume $k\ll aH$ when
we discuss the evolution of the curvature perturbation around the
overshooting regime. In the following discussion we denote the
coefficient of the second and third terms in eq.(\ref{superevo}) by
$A$ and $B$, respectively; their dependence on $N$ is depicted in
Fig.~\ref{ab}. After horizon crossing, the evolution of $\zeta_k$
undergoes four stages, which are shown in Fig.~\ref{reimk11}:

\begin{figure}[h!]
\centering
\subfloat[$31< N <34$]{\includegraphics[width=70mm]{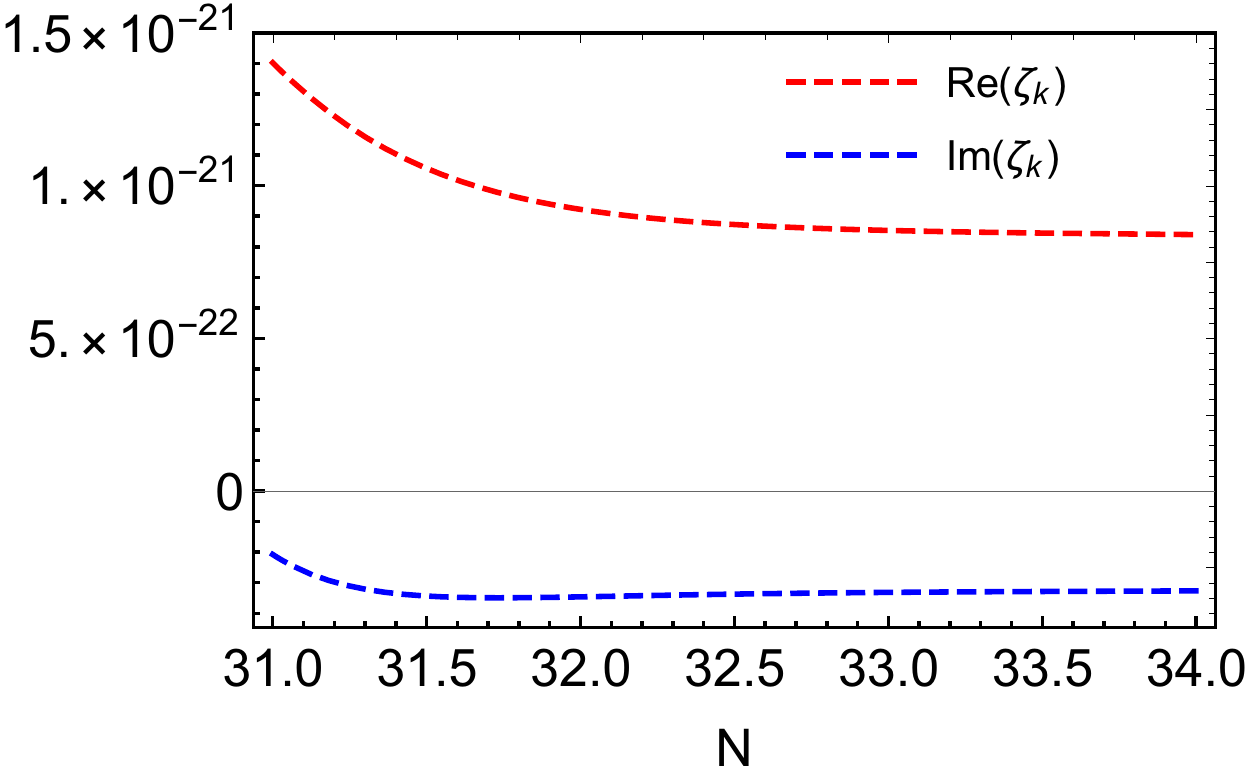}
\label{reimk11a}	}
\subfloat[$34< N < 37$]{\includegraphics[width=70mm]{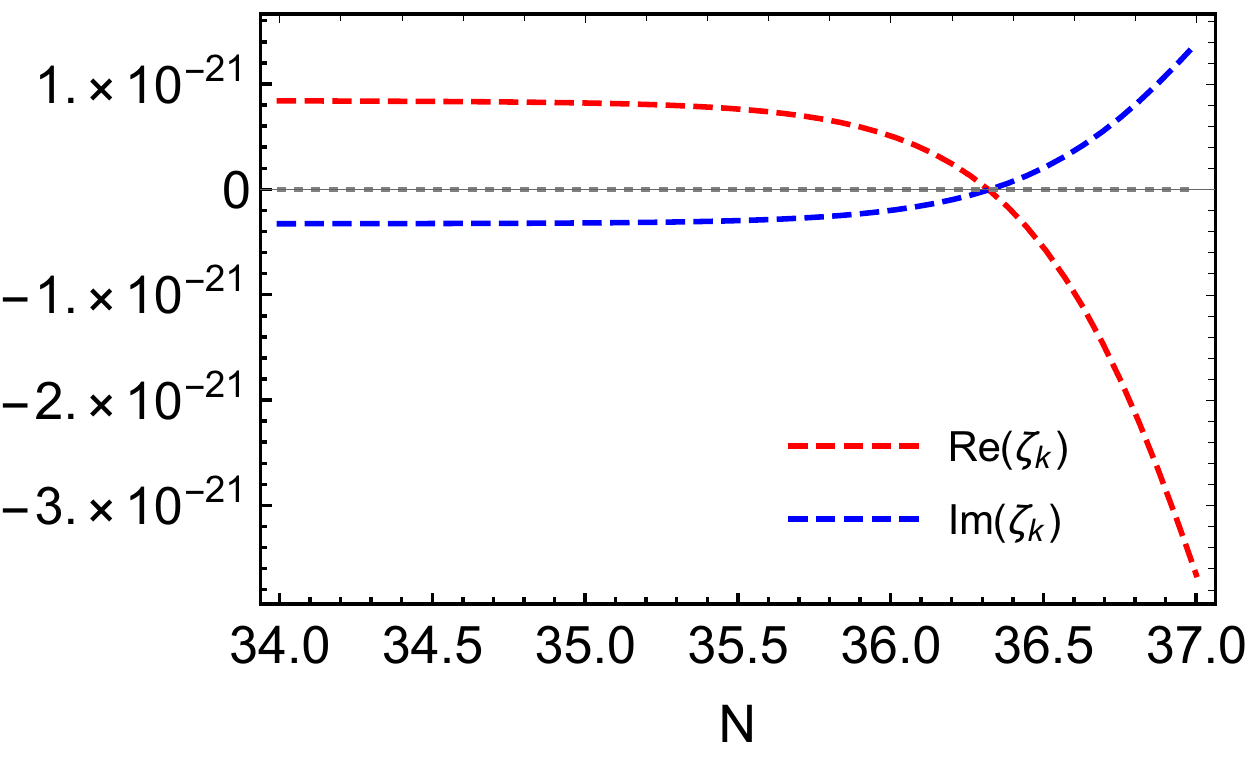}
\label{reimk11b}	}
\hspace{0mm}
\subfloat[$37< N <38$]{\includegraphics[width=70mm]{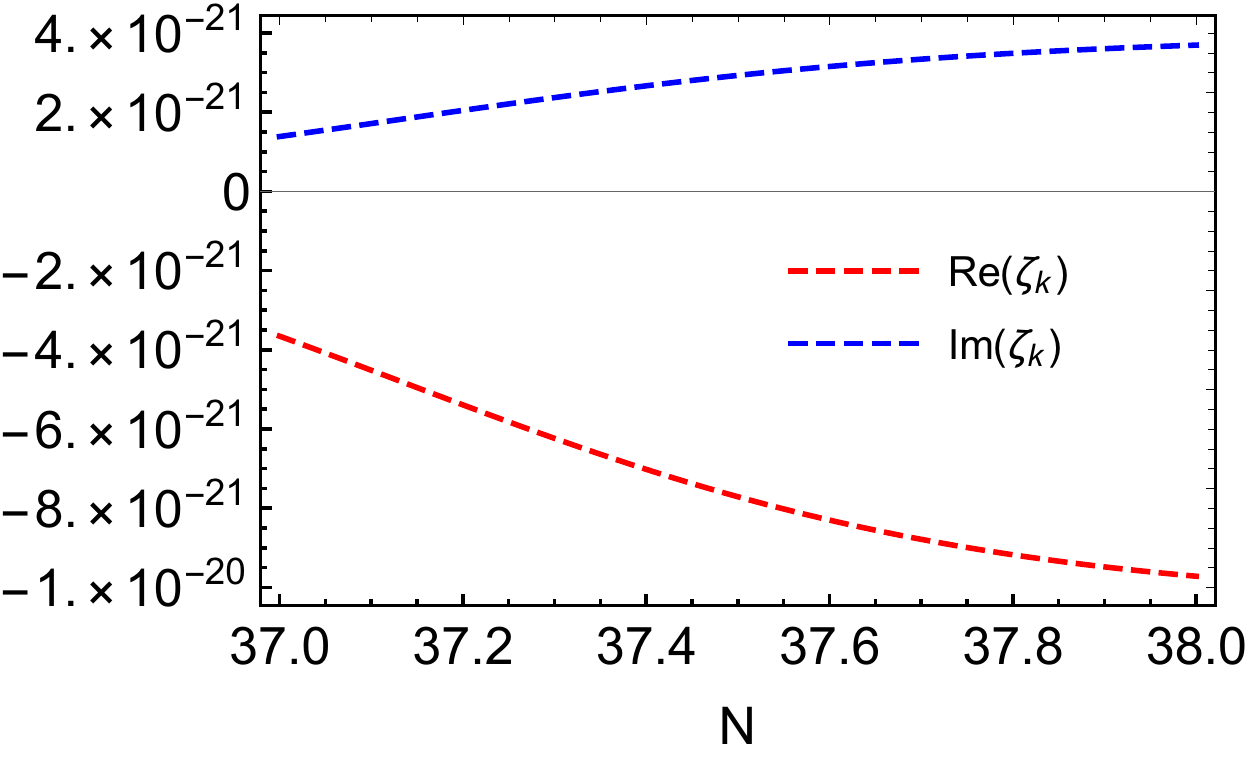}
\label{reimk11c}	}
\subfloat[$38< N <70$]{\includegraphics[width=70mm]{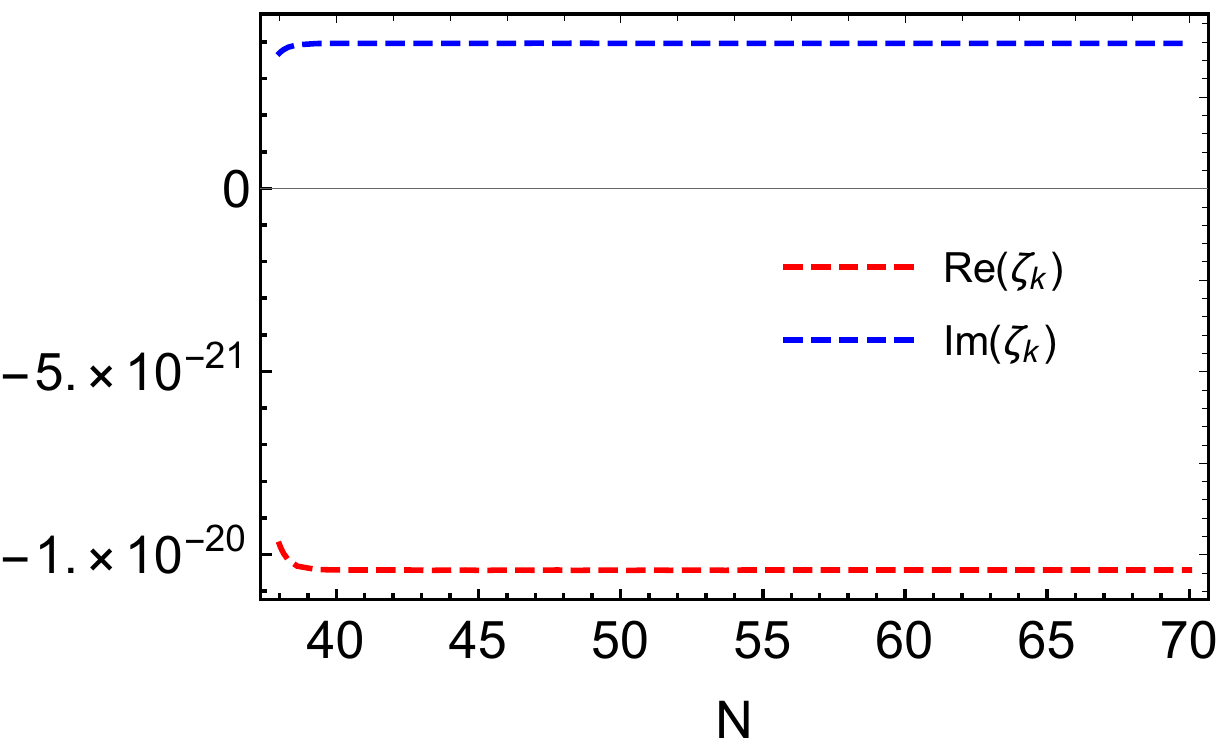}
\label{reimk11d}	}
\caption{Evolution of curvature perturbations for the mode with
  co--moving wave number $k= 10^{11} \ \mathrm{Mpc^{-1}} $.}
\label{reimk11}
\end{figure}

\begin{itemize}
	
\item $31< N <34$, Fig.~\ref{reimk11a}: in this region, $A$ is always
  positive and therefore acts as a friction term, and $B$ decreases
  exponentially. Since we are already beyond horizon crossing,
  $B^2 - A^2/4 < 0$, i.e. eq.(\ref{superevo}) approximately describes
  an over--damped oscillator. This means that $\zeta_k$ does not
  oscillate any more, but decreases rather slowly. This explains the
  first, short flat region in Fig.~\ref{plotk11}. In this region the
  second derivative of $\zeta_k$ can be neglected, thus the curvature
  perturbation satisfies
  $\frac{d\zeta_k}{d N} \approx \frac{B}{A}\zeta_k$. This
  approximation ceases to hold somewhat before the value $N_0$ where
  $A$ turns to zero, i.e. where the overshooting region starts; recall
  that in our case $N_0 \approx 34$.
	
\item $34< N < 37$, Fig.~\ref{reimk11b}: for $N \geq N_0$,
  $B \leq 0.01$ (see Fig.~\ref{ab}) has become essentially negligible,
  while $A$ changes from positive to negative hence acts as a driving
  term. As discussed above this leads to an exponential increase of
  the first derivative of $\zeta_k$. Since the epoch of exponentially
  decreasing first derivatives is considerably shorter than for
  $k = 2.5 \cdot 10^{10} \ \mathrm{Mpc^{-1}}$, $\zeta_k$ itself now
  begins to vary appreciably already at $N \approx 36$.
	
  Remarkably, shortly thereafter both the real and imaginary parts
  cross the zero point nearly at the same time, leading to
  $\zeta_k \to 0$. This can be understood from the approximate
  solution for $\zeta_k$ in this range of $N$:
\begin{equation} \label{zetasolution}
\zeta_k (N) \approx \zeta_k (N_0) - \frac{1}{A} \frac{d\zeta_k}{dN}
\Big|_{N_0} \left({\rm e}^{-A(N-N_0)}-1\right).
\end{equation}
Since $\frac{d\zeta_k}{dN} \Big|_{N_0} \propto \zeta_k (N_0)$, we see
that the real and imaginary parts of $\zeta_k(N)$ go through zero at
the same point, which is also the origin of the very sharp minimum
depicted in Fig.~\ref{plotk11}. Note that for the previous case,
$k = 2.5 \cdot 10^{10} \ \mathrm{Mpc^{-1}}$, the overshooting epoch
ended before $\zeta_k$ reached zero; the sharp minimum of the final
power spectrum depicted in Fig.~\ref{ms} corresponds to that value of
$k$ where the overshooting epoch lasts just long enough to drive
$\zeta_k$ to zero, and then ends. In the case at hand instead
$|\zeta_k|$ again increases exponentially beyond the zero crossing.
	
\item $37< N <38$, Fig.~\ref{reimk11c}: this is the stage just after
  the overshooting epoch, i.e. $A$ is again positive so that the
  modulus of the first derivative of $\zeta_k$ is decreasing
  exponentially again. For a while $|\zeta_k|$ keeps increasing,
  albeit more slowly than before. Of course, $B \leq 10^{-5}$ is now
  completely negligible.
	
\item $38< N <70$, Fig.~\ref{reimk11d}: the overshooting phase has
  ended and the universe comes back to (U)SR inflation, therefore
  curvature perturbations are frozen again at the super horizon scale
  and matter perturbation evolve adiabatically again. This corresponds
  the second flat region shown in Fig.~\ref{plotk11}. For this value
  of $k$ the second flat region is already higher than the first one,
  thus the analytical SR approximation underestimates the final power.
\end{itemize}

\begin{figure}[h!]
\centering
\subfloat[]{\includegraphics[width=70mm]{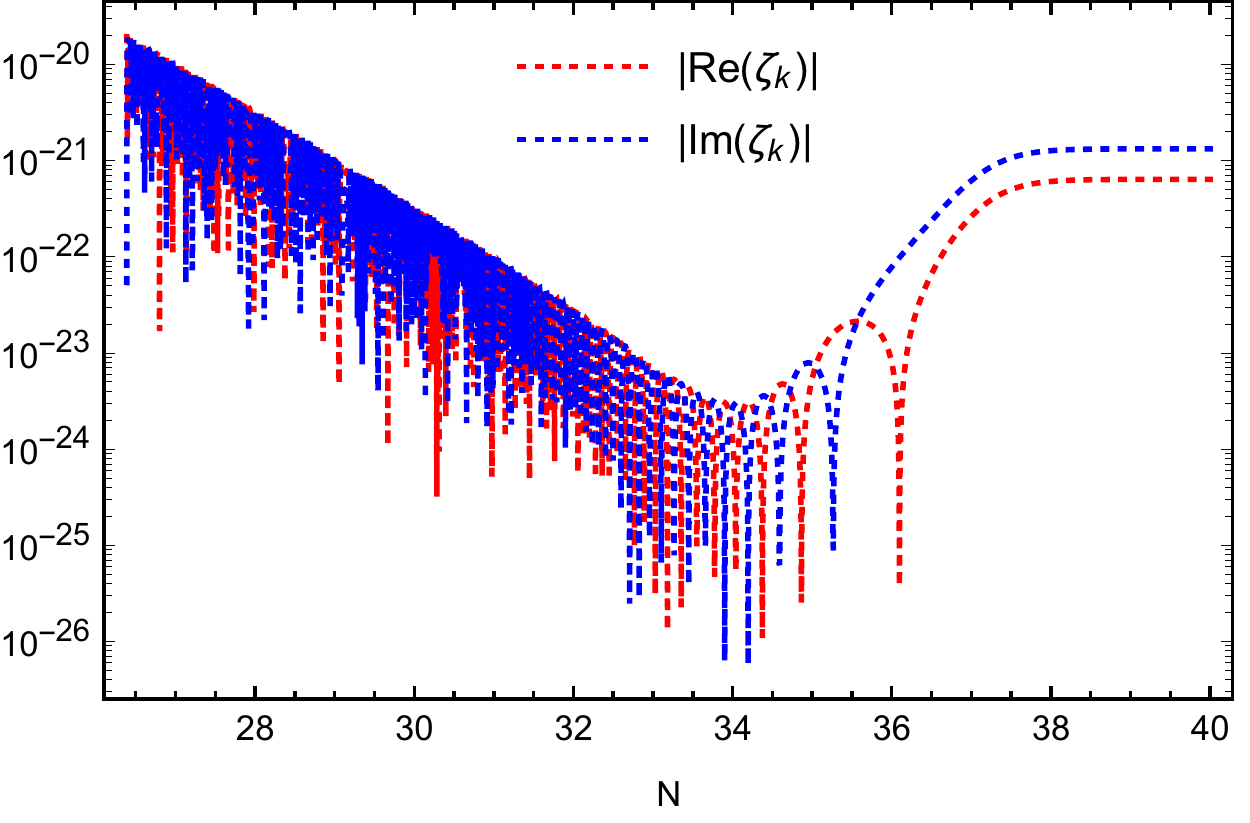}	\label{xi_k13}	}
\subfloat[]{\includegraphics[width=70mm]{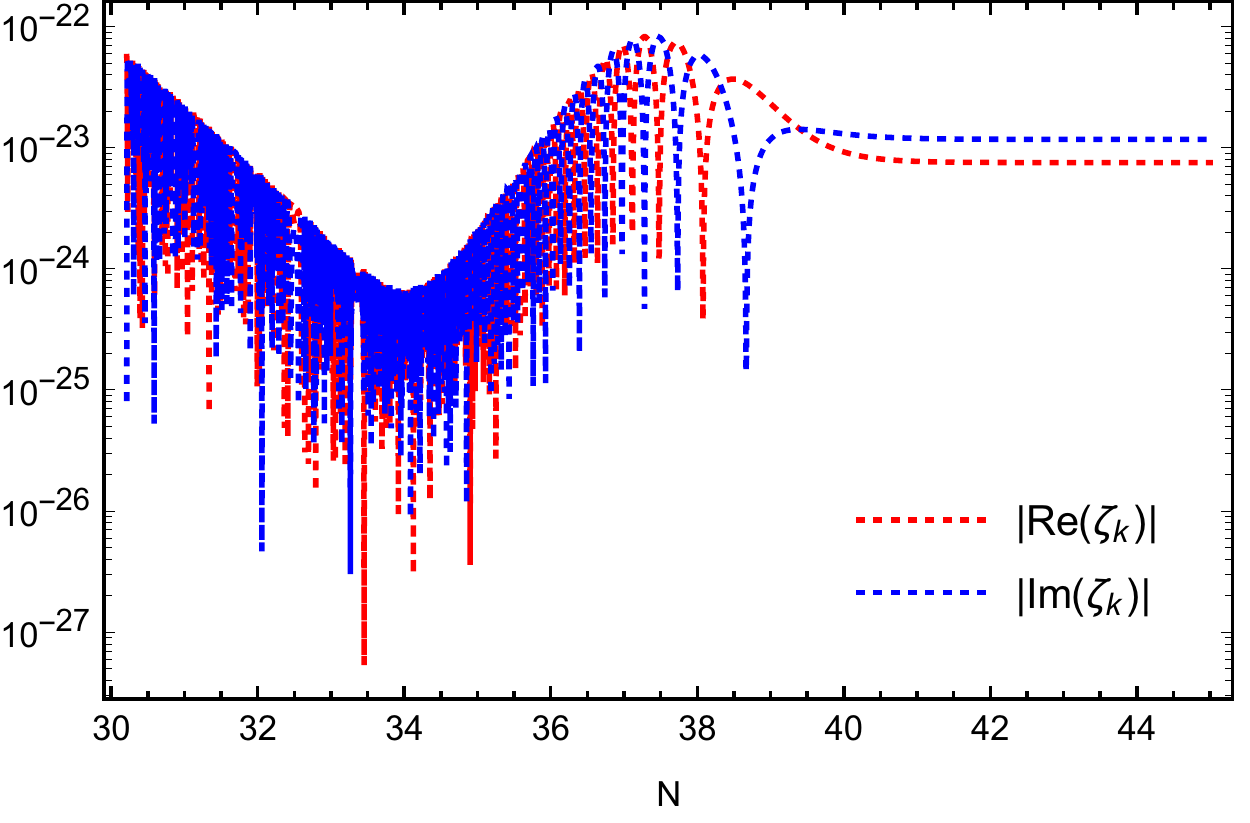}	\label{xi_k14}	}
\caption {Evolution of curvature perturbations for the modes with
  co--moving wave number $k= 10^{13} \ \mathrm{Mpc^{-1}}$ (left) and
  $k= 2 \times 10^{14} \ \mathrm{Mpc^{-1}}$ (right).}
\end{figure}
%\FloatBarrier

Fig.~\ref{plotk13}: for the mode with
$k=10^{13} ~ \mathrm{Mpc^{-1}}$, horizon crossing occurs at
$N = 36.5$; this lies in the middle of the overshooting regime. Hence
there is no overdamped oscillator phase, and therefore also no plateau
in the evolution of $\mathcal{P}_\zeta$, before the overshooting
epoch, unlike in Figs.~\ref{plotk10} and \ref{plotk11}.

Nevertheless for $N < 34$ eq.(\ref{superevo}) again describes a damped
oscillator, the amplitude of the oscillation decreasing
$\propto {\rm e}^{-N}$. This is also shown in
Fig.~\ref{xi_k13}. However, at $N = N_0 \simeq 34$ the second term in
eq.(\ref{superevo}) changes sign, eventually reaching $-3$ as shown in
Fig.~\ref{ab}. For $N_0 < N < N_{\rm cross}$ eq.(\ref{superevo})
therefore leads to oscillations whose amplitude {\em grows}
$\propto {\rm e}^{2N}$.  For $N > N_{\rm cross}$ the approximately
exponential growth continues for a while, this time for $|\zeta_k|$
itself which no longer oscillates. As before, for $N > 37$ the
derivative of $\zeta_k$ begins to decrease exponentially in magnitude,
which leads to $\zeta_k$ itself becoming essentially constant for
$N > 38$, after the end of the overshooting epoch. Not surprisingly,
for this mode the analytical SR estimate for the final power also
fails.

Fig.~\ref{plotk14}: for the mode with co--moving wave number
$k=2\times 10^{14} ~ \mathrm{Mpc^{-1}}$, horizon crossing takes place
at $N = 39.5$, i.e. during USR well after the end of the overshooting
epoch. We again see an (initially very rapid) oscillation whose amplitude
first drops $\propto {\rm e}^{-N}$ and then increases $\propto {\rm e}^{2N}$
once $N > 34$, i.e. in the overshooting regime. Since for this value of
$k$ the overshooting epoch ends before horizon crossing, for $N > 38$
the function $\zeta_k$ again undergoes a few oscillations with exponentially
decreasing amplitude, before settling into an overdamped oscillator mode,
i.e. approaching a constant.

In this case the analytical SR approximation for the final power
actually works quite well. On the one hand this may not be surprising,
since the arguments of Sec.~2 imply that perturbations are now
adiabatic at super--horizon scales. On the other hand, it may be
surprising that one still gets the correct result by imposing the
initial conditions (\ref{initialconds}) just a couple of e--folds before
horizon crossing. This implies that these initial conditions capture
the dynamics of the MS equation even in the overshooting regime, as
long as the mode is still (deep) inside the horizon.

In fact, eq.(\ref{initialconds}) shows that $|v_k|$ is simply a
constant (independent of $N$) for sub--horizon modes. The dynamics is
therefore entirely captured by the factor $1/z$ which relates
$\zeta_k$ to $v_k$, see eq.(\ref{zdef}). This contains a factor
$1/a \propto {\rm e}^{-N}$, which dominates the $N-$dependence in the
(U)SR regime where the SR parameter $\epsilon_H$ is approximately
constant (and small). However, eq.(\ref{newreviseddn}) shows that in
the overshooting regime, where the term containing the derivative of
the potential can be neglected,
$\sqrt{\epsilon_H} \propto |d \chi / dN| \propto {\rm e}^{-3N}$, so
that altogether
$|\zeta_k| \propto 1/(a \sqrt{\epsilon_H}) \propto {\rm e}^{+2N}$, as
we had inferred from the MS equation.

For modes with even larger $k$, the situation is very similar to the
case with $k=2\times 10^{14} ~ \mathrm{Mpc^{-1}}$, i.e. the analytical
SR approximation for the final power agrees with the numerical result,
as shown in Fig.~\ref{ms}, as long as the modes cross out of the
horizon (well) before the end of inflation. The approximation fails
again for modes with very large $k$ which cross the horizon near the
end of inflation where SR again fails; however, we know of no way to
probe those modes observationally.

From the above discussion it is easy to understand that the peak in
the power spectrum shown in Fig.~\ref{ms} occurs for the mode which
crosses the horizon just at the beginning of the overshooting
regime. In this case, $N_{\rm SR}=0$, and the exponential increase of
$|d\zeta_k/dN|$ is maximized. This also greatly enhances the final
value of $|\zeta_k|$, leading to the maximum in the spectrum. We find
the maximal scalar power spectrum is
$\mathcal{P}_{\zeta} \approx 1.1 \times 10^{-4}$ for
$k = 1.1 \cdot 10^{13} \mathrm{Mpc^{-1}}$. According to
ref.~\cite{Ezquiaga:2017fvi}, fluctuations of this size are large
enough to lead to significant formation of primordial black holes,
which might even constitute a sizable fraction of all dark matter. In
the next section we will investigate another cosmological consequence
of such a large curvature perturbation, namely the amplification of
primordial gravitational wave signatures due to second order effects.

Before closing this section we comment on possibilities to increase
the power spectrum even further. According to
ref.~\cite{Ezquiaga:2018gbw}, quantum diffusion effects can in
principle further enhance the power; however, we checked that in our
case one always has $|\dot{\chi}| \gg H^2/(2\pi)$, which indicates
that quantum diffusion does not change the evolution of the inflaton
field significantly. Moreover, refs.~\cite{Ballesteros:2017fsr,
	Rasanen:2018fom} argue that a shallow local minimum of the inflaton
potential can also enhance the power spectrum; this agrees with our
finding in eq.(\ref{curvature_evolution}), because the curvature of
the potential $V^{\prime \prime}$ is maximal near a local
minimum. However, the inflaton might get stuck in a local minimum, in
which case inflation would never end. In contrast, the scenario we
presented leads to a well--behaved inflationary epoch, in agreement with
current observations.

\section{Second Order Gravitational Wave  Signatures}
\label{gwsig}

As well known, primordial perturbations of the inflaton field source
primordial gravitation waves. Usually the strength of the GW signal is
estimated in linear order in perturbations; for SR inflation, this
leads to the famous prediction $r = 16 \epsilon_V$, where $r$ is the
tensor--to--scalar ratio. However, in some cases effects that are
second order in the curvature perturbations can also contribute
significantly to the primordial GW signal \cite{Matarrese:1997ay,
  Mollerach:2003nq, Baumann:2007zm,Martineau:2007dj}. As has recently
been emphasized in \cite{Gong:2017qlj}, which analyses a polynomial
potential with an inflection point, this occurs in particular when an
overshooting regime enhances the power spectrum. In the following
analysis, we mainly follow the formalism given in
\cite{Baumann:2007zm, Gong:2017qlj}.

In the radiation era, the second order tensor perturbation with
comoving wave number $k$ satisfies \cite{Ananda:2006af,
	Baumann:2007zm, Garcia-Bellido:2017aan, Saito:2009jt, Bugaev:2009zh,
	Bugaev:2010bb, Alabidi:2012ex, Gong:2017qlj}:
\begin{equation} \label{gweq}
h^{\prime \prime }(\vec{k},\tau) + 2 a H h^{\prime}(\vec{k},\tau) +
k^2 h(\vec{k},\tau) = S(\vec{k},\tau)\,,
\end{equation}
where a prime denotes a derivative with respect to the conformal time
$\tau$. $S(\vec{k},\tau)$ denotes the source term, which is given by
\cite{Gong:2017qlj}
\begin{equation} \label{source}
\begin{split}
S(\vec{k},\tau) = \int \frac {d^3 \tilde{k}} {(2\pi)^{3/2}} \tilde{k}^2
\Bigg[ 1 &- \Bigg( \frac {\vec{k} \cdot  \vec{\tilde k} } {k \tilde{k}}
\Bigg)^2 \Bigg] \Bigg[ 12 \Phi ( \vec{k} - \vec{\tilde k},\tau)
\Phi (\vec{\tilde k},\tau) \\
&+ 8 \Bigg( \tau \Phi (\vec{k} - \vec{\tilde k},\tau)
+ \frac {\tau^2} {2} \frac {d \Phi (\vec{k} -\vec{\tilde k},\tau) } {d\tau}
\Bigg) \frac {d \Phi (\vec{\tilde k},\tau) } {d \tau} \Bigg]\,.
\end{split}
\end{equation}
The Bardeen potential appearing in eq.(\ref{source}) is related to the
curvature perturbation via $\Phi = 2\zeta_k/3$ \cite{Gong:2017qlj}. As
we have seen in the last section, the scalar curvature perturbation is
enhanced during an overshooting regime, thus we expect that the source
term for gravitational waves will also be enhanced.

In order to obtain the current gravitational wave density, we have to
solve eq.(\ref{gweq}) with source given by eq.(\ref{source}). To that
end we'll apply the Green's function method of
ref.~\cite{Baumann:2007zm}. Rewriting eq.(\ref{gweq}) with $v:= a h$,
we get
\begin{equation} \label{veq}
v^{\prime \prime }(\vec{k},\tau) +\left( k^2 - \frac {a^{\prime \prime }}{a} \right)
v(\vec{k},\tau) = a S(\vec{k},\tau)\,.
\end{equation}
The solution of eq.(\ref{gweq}) can then be written as
\begin{equation}
h(\vec{k},\tau) = \frac {1} {a(\tau)} \int d\tilde{\tau}
g(\vec{k}, \tau; \tilde \tau ) \left[ a(\tilde \tau) S(\vec{k},\tau)
\right]\,,
\end{equation}
where $g$ is the Green's function for eq.(\ref{veq}), which satisfies:
\begin{equation}
g^{\prime \prime } ( \vec{k}, \tau; \tilde\tau ) + \left(
k^2- \frac {a^{\prime \prime }} {a} \right)
g( \vec{k}, \tau; \tilde\tau ) = \delta (\tau -\tilde{\tau})\,.
\end{equation}
Once the tensor perturbations are known, we can further compute the
contribution of these primordial gravitational waves to the total
energy budget of the universe.  For a matter-dominated universe, one has  \cite{Baumann:2007zm}:
\begin{equation} \label{secondgw}
\Omega^{(2)}_{\rm GW}(k,\tau) = A^{(2)}_{\rm GW} \mathcal{P}^2_{\zeta} \cdot
\begin{cases}
\frac {a(\tau)} {a_{\rm eq}} \frac {k} {k_{\rm eq}} & \text{if}\ k< k_{\rm eq} \\
\frac {a(\tau)} {a_{\rm eq}} \left( \frac {k} {k_{\rm eq}} \right)^{2-2\gamma}&
\text{if}\ k_{\rm eq}< k< k_c(\tau)\\
\frac {a_{\rm eq}} {a(\tau)} & \text{if}\ k> k_c(\tau)\\
\end{cases} \, .
\end{equation}
Here
$k_c(\tau) = \left( \frac {a(\tau)} {a_{\rm eq}}
\right)^{1/(\gamma-1)} k_{\rm eq}$, $ A^{(2)}_{\rm GW} \simeq 10$, and
$\gamma \simeq 3$ \cite{Baumann:2007zm}. Finally,
$k_{\rm eq} \approx 0.01\mathrm{Mpc^{-1}}$ \cite{Aghanim:2018eyx}
denotes the wave number that re--entered the horizon when matter and
radiation had the same energy density and
$a_{\rm eq} \equiv 1/(1+z_{\rm eq})$ denotes
the scale factor at that time. Eqs.(\ref{secondgw}) hold after
matter--radiation equilibrium, i.e. for $\tau > \tau_{\rm eq}$ where
$k_c(\tau) > k_{\rm eq}$. We are interested in the gravitational wave
signatures in the range of wave numbers that are enhanced by the
scalar perturbation in the overshooting regime,
$10^{11} \mathrm{Mpc^{-1}} < k< 10^{14} \mathrm{Mpc^{-1}}$ (see
Fig.~\ref{ms}). These are much larger than $k_c(\tau_0)$; the present
($\tau = \tau_0$ with $a(\tau_0) = 1$) GW signal is thus
\cite{Baumann:2007zm, Gong:2017qlj} 
\begin{equation} \label{final_gw}
\Omega^{(2)}_{\rm GW}(k,\tau_0) \simeq 10 \  \mathcal{P}^2_{\zeta} \ a_{\rm eq}\,.
\end{equation}

Using $z_{eq} =3387$, $H_0 = 67.4 \ \mathrm{km \ s^{-1} Mpc^{-1}}$
\cite{Aghanim:2018eyx} and the power spectrum $\mathcal{P}_{\zeta}$
computed via the MS formalism in the last section, we can calculate
the current gravitational wave energy density due to this second order
effect.
\begin{figure}[h!]
	\centering
	\includegraphics[width=.6\paperwidth, keepaspectratio]{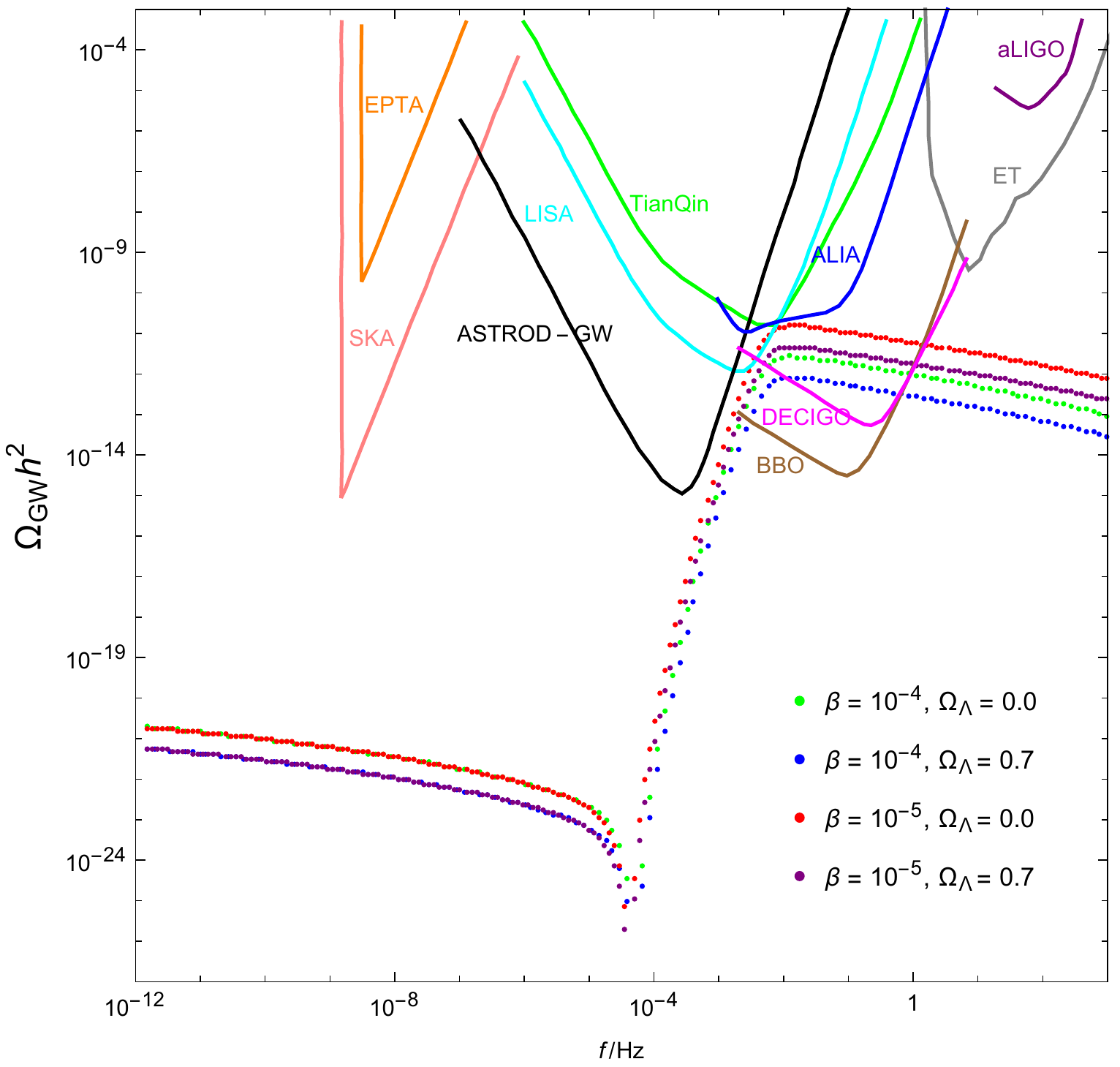}
	\caption {The gravitational wave signal from CHI induced by second
		order effects, $\Omega_{\rm GW} \propto \mathcal{P}_\zeta^2$. In the
		above plot, $f = \frac{c k}{2 \pi}$ is the frequency of the
		gravitational wave, with $c$ the speed of light. The frequency range
		we have shown is $10^{-12} ~\mathrm{Hz}< f < 10^3~\mathrm{Hz}$,
		which corresponds to
		$10^{3} ~ \mathrm{Mpc^{-1}} < k <6.5 \times 10^{17}
		~\mathrm{Mpc^{-1}}$. The experimental sensitivity curves we show
		include the Square Kilometer Array (SKA),
		Einstein Telescope (ET), Astrodynamical Space Test of Relativity
		using Optical-GW detector (ASTROD-GW), Advanced LIGO (aLIGO),
		European Pulsar Timing Array (EPTA), Laser Interferometer Space
		Antenna (LISA) \cite{lisa}, Advanced Laser Interferometer Antenna
		(ALIA), Big Bang Observer (BBO), Deci-hertz Interferometer GW
		Observatory (DECIGO) and TianQin \cite{Luo:2015ght}. The
		sensitivities of EPTA, SKA, LISA, TianQin and aLIGO limit are taken
		from \cite{Gong:2017qlj}. The ALIA, DECIGO, and BBO sensitivity
		curves are reproduced from \cite{Moore:2014lga}. The ASTROD-GW and
		ET curves are adapted from \cite{Kuroda:2015owv}. }
	\label{gwden}
\end{figure}

The result is shown in Fig.~\ref{gwden}, which also shows the
sensitivity of several planned gravitational wave detectors.  We saw
at the end of the last Section that the maximum of the power spectrum
is at $k = 1.1 \times 10^{13} \mathrm{Mpc^{-1}}$, which corresponds to
frequency $f = 0.017$ Hz. This is near the frequency of maximal
sensitivity of the upcoming space mission LISA, which may just barely
be able to detect this signal if the parameter $\beta = 10^{-5}$ (red
curve), while for $\beta = 10^{-4}$ (green) the signal is below the
foreseen LISA sensitivity. Recall that the CMB predictions for both
values of $\beta$ are consistent with latest Planck measurements, see
Sec.~\ref{cmbprediction}. Since a larger $\beta$ makes the potential
less flat in the USR region and reduces $V^{\prime \prime}$ in the
overshooting region, it is expected that the corresponding curvature
perturbation is less enhanced compared to that with a smaller $\beta$
according to eq.(\ref{curvature_evolution}). This explains why the
peak of GW signatures with $\beta = 10^{-4}$ is lower. However, the
second generation space missions DECIGO and BBO should easily detect
this signal even for $\beta =10^{-4}$.

As already noted, eqs.(\ref{secondgw}) hold for a matter-dominated universe, i.e. for $\tau > \tau_{\rm eq}$ where $k_c(\tau) > k_{\rm eq}$. We are not aware of a calculation of $\Omega_{\rm GW}^{(2)}$ that includes dynamical effects due to a cosmological constant; this would be required for very large wavelengths, which crossed the horizon after the cosmological constant (or, more generally, dark energy) contributed significantly to the total energy density. Fortunately we are interested in much shorter wavelengths, with $10^{11}$ Mpc$^{-1} < k < 10^{14}$ Mpc$^{-1}$, which crossed the horizon when the cosmological constant was entirely negligible. The further dilution of the gravitation wave signal by the recent accelerated expansion of the universe can then simply be described by multiplying the right-hand side of eqs.(\ref{secondgw}) with the normalized matter density $\Omega_m(\tau)$, yielding a suppression by a factor $\simeq 0.3$ today.\footnote{We thank the anonymous referee for guiding us to this interesting effect of the cosmological constant.} These results are shown by the purple and blue lines respectively for $\beta = 10^{-5}$  and $\beta = 10^{-4}$.

Fig.~\ref{gwden} also shows that the peak of the CHI signal lies at
frequencies that are too large for the pulsar timing arrays even after
SKA comes on--line. The size of the signal is well below the
sensitivity of advanced LIGO, and even below that of the planned
Einstein Telescope (ET).

\section{Summary and Conclusions}
\label{conclusion}

In this paper, we have revisited critical Higgs inflation, carefully
computing the power spectrum as well as the gravitational wave
signatures induced by second order effects.

In Sec.~2 we analyzed the evolution of curvature perturbations under
(ultra--)slow roll as well as overshooting conditions in general
terms. In the former, the second derivative of the inflaton field with
respect to time can be neglected in the equation of motion; we showed
that the perturbations are adiabatic in this case, which further
implies that the curvature perturbations are frozen at super--horizon
scales. This allows one to calculate the final power spectrum (at the
end of inflation), which seeds all observed structures in the
universe, by simply computing the power spectrum at horizon
crossing. We emphasize that this also holds for ultra--slow roll
(USR), which in our model describes the epoch when the inflaton field
is near the (almost) saddle point of the potential. Here the
deviations from adiabacity are even smaller than in the SR case, so
that the usual approximate treatment is even more accurate.

In contrast, when the inflaton enters an overshooting phase where the
acceleration $|\ddot\chi|$ is much larger than the derivative of the
potential $|V^{\prime}|$, we showed that perturbations are no longer
adiabatic; this can be described in terms of entropic pressure
perturbations. In this case the curvature perturbations are {\em not}
conserved at super--horizon scales, so that the standard SR
approximation for calculating the power spectrum is expected to break
down. To our knowledge this is the first time that the significance of
entropic perturbations has been discussed in this context. Our
eq.(\ref{curvature_evolution}) shows that the enhancement of the
perturbations after horizon crossing but during the overshooting epoch
will increase for larger curvature $V^{\prime \prime}$ of the
potential. This can be very useful for inflationary model building if
one wants to strongly enhance the power spectrum, e.g. in order to
produce primordial black holes. See \cite{Fu:2020lob} for a recent investigation along this direction. 

In Sec.~3 we illustrated these general results by analyzing the CHI
scenario in detail. For judiciously chosen parameters, an overshooting
epoch appears between the SR and USR eras. During the overshooting
stage the Hubble SR parameters vary rapidly, which implies the
universe deviates significantly from SR evolution. As a result the
usual analytical approximation to compute the power spectrum fails. We
instead solved the Mukhanov--Sasaki equation numerically to compute
the power spectrum at the end of inflation.  We find that the modes
which cross the horizon just before or during the overshooting epoch
are greatly enhanced. The power spectrum can reach values of order
$\sim 10^{-4}$ for $k \sim 10^{13}\ \mathrm{Mpc^{-1}}$; this is to be
compared to values of order $10^{-9}$ at the (much smaller) $k-$values
probed by the CMB anisotropies. These results differ quantitatively
from those of ref.\cite{Ezquiaga:2017fvi}, where the power spectrum
was computed in the SR approximation.

In the course of this discussion we found a version of the MS equation
very useful which holds for the $k-$space perturbation $\zeta_k$
directly, rather than for the related quantity $v_k$ which is usually
employed, see eq.(\ref{superevo}). This allowed us to understand the
numerical results in detail: why the SR approximation agrees with the
MS formalism for modes that cross out of the horizon well before
(small $k$) or well after (large $k$) the overshooting epoch; why
there is a sharp minimum in the power spectrum, for scales that cross
the horizon a few e--folds before the overshooting region; and where
the maximum of the spectrum lies. These findings are generic and can
also be applied to explain the numerical power spectrum results for
other inflation models featuring a near--inflection point, for example
\cite{Gong:2017qlj, Ezquiaga:2018gbw, Cheng:2018qof}, where detailed
explanations concerning the numerical results are not given. 

Finally we analyzed the second order GW signatures induced by the
enhanced scalar perturbations. The strength of this signal is
proportional to the square of the scalar power spectrum. The peak of
the latter at co--moving wave number of order
$10^{13} \ \mathrm{Mpc^{-1}}$ corresponds to a peak of the GW signal
at a frequency of $0.017$ Hz. We find that for our choices of
parameters, the GW signal should remain detectable up to frequency of
order $1$ Hz by two planned second--generation space based GW
experiments, DECIGO and BBO. Detection of this signal is a firm
prediction, if the power spectrum is enhanced to the level that might
allow significant production of PBHs. This statement holds also in
other models of inflation proposed recently
\cite{Garcia-Bellido:2017mdw, Gao:2018pvq, Dalianis:2018frf,
  Ozsoy:2018flq, Cicoli:2018asa, Hertzberg:2017dkh,
  Ballesteros:2017fsr, Gong:2017qlj, Kannike:2017bxn,
  Cheng:2018qof}. Hence if future GW experiments fail to detect this
signal, one could conclude that no significant PBH formation occurred
immediately after inflation.

In this paper we did not consider effects due to non--Gaussianity.
Since PBHs only form in regions with large overdensity, their
formation rate can be greatly enhanced if there are significant
non--Gaussian tails in the distribution function of the density
perturbations \cite{Franciolini:2018vbk, Atal:2018neu}. It should be
noted that the calculation of the PBH formation rate is in any case
somewhat uncertain; however, the second order GW signal computed in
the Gaussian approximation should be detectable by second generation
space missions for the entire range of perturbations that could
plausibly lead to sizable PBH formation, unless non--Gaussianities are
quite large for the relevant modes. In this context it is important to
note that according to a recent analysis
\cite{Cai:2018dig,Unal:2018yaa} primordial non--Gaussianities will
also enhance the second order GW signal itself. We leave a detailed
investigation of the impact on non--Gaussianities on CHI inflation for
future work.

\section*{Acknowledgments}

We are grateful to Guillermo Ballesteros for helpful correspondence
regarding the numerical solution of the Mukhanov--Sasaki equation in
\cite{Ballesteros:2017fsr}. We also thank Fazlollah Hajkarim and
George Tringas for discussions, and Jose Mar\'ia Ezquiaga for email
communications concerning \cite{Ezquiaga:2018gbw}.

\appendix
\section{Mukhanov-Sasaki Equation and Its Analytical Solution }
\label{solums}

In this Appendix we briefly review the derivation of the MS equation and
discuss its analytical solution in (quasi) de Sitter spacetime.

\subsection{Derivation of the Mukhanov-Sasaki Equation}

We start from the general action of a single real scalar field
minimally coupled to gravity,
\begin{equation} \label{scalaraction}
S = \frac {1} {2} \int d^4x \sqrt{-g} \left[ R - g^{\mu \nu} \triangledown_{\mu}
\phi \triangledown_{\nu} \phi - 2V(\phi) \right]\,.
\end{equation}
Using the Arnowitt--Deser--Misner (ADM) formalism
\cite{Arnowitt:1962hi} this action can be expanded as:
$S= S_{(0)} + S_{(1)} + S_{(2)}+ \dots$, where the order is with
respect to the perturbation $\zeta$. $S_{(0)}$ denotes the background,
$S_{(1)}$ vanishes due to the first order Hamiltonian constraint
equation \cite{Maldacena:2002vr, Seery:2005wm, Chen:2006nt}, $S_{(2)}$
contains the two--point correlation function we wish to compute, and
higher orders contribute to non--Gaussian contributions to the power
spectrum which are beyond the scope of our analysis. In
\cite{Maldacena:2002vr, Chen:2006nt}, it is shown that the action up
to the second order of $\zeta$ can be written as
\begin{equation} \label{secorderact}
S_{(2)} = \frac {1} {2} \int d^4x a^3 \frac {\dot{\phi}^2} {H^2} \left[
\dot{\zeta}^2 - a^{-2}(\partial_i\zeta)^2 \right] \,;
\end{equation}
here $a$ is the scale factor in the FRW metric. Now define the
Mukhanov variable as
\begin{equation} \label{vzr}
v\equiv -z \zeta \,,
\end{equation}
where $z$ carries the information about the background field:
\begin{equation} \label{za}
z^2 \equiv a^2 \frac {\dot{\phi}^2} {H^2} = 2 a^2 \epsilon_H\,.
\end{equation}
%
%Using the spatially flat gauge, i.e. $\Psi=0$ and
%$\zeta = H \delta \phi /\dot{\phi}$, we can relate the Mukhanov
%variable $v$ to the quantum fluctuation $\delta \phi$:
%
%\begin{equation} \label{vadeltaphi}
%v = a \delta \phi.
%\end{equation}
%

Transforming the cosmic time $t$ to the conformal time $\tau$ with
$d\tau = dt/a$, we can rewrite the action eq.(\ref{secorderact}) as
\begin{equation} \label{secorderact2}
\begin{split}
S_{(2)} & = \frac {1} {2} \int d\tau d^3x
\left[ (v^{\prime})^2 - (\partial_iv)^2 + \frac {z^{\prime \prime}}{z} v^2 \right]\\
&= \frac {1} {2} \int d\tau d^3x \mathcal{L}_{(2)}\,.
\end{split}
\end{equation}
Here a prime denotes a derivative with respect to $\tau$. The
Euler--Lagrange equation derived from $\mathcal{L}_{(2)}$ reads
\begin{equation} \label{mukhanovdiff}
\frac {\partial \mathcal{L}_{(2)}} {\partial v}
- \frac {\partial} {\partial \tau } \left( \frac{\partial \mathcal{L}_{(2)}}
{\partial v^{\prime}} \right)
- \frac {\partial} {\partial x^{i} } \left(
\frac {\partial \mathcal{L}_{(2)}} {\partial_i v} \right) =0\,.
\end{equation}
Plugging $\mathcal{L}_{(2)}$ from eq.(\ref{secorderact2}) into
eq.(\ref{mukhanovdiff}), we obtain:
\begin{equation} \label{mukhanovdiff2}
v \frac{z^{\prime \prime}}{z} -v^{\prime \prime} + \partial^i \partial_i v =0.
\end{equation}

It is usually more convenient to analyse the perturbations in Fourier
space. To that end we write the field $v$ as:
\begin{equation} \label{fourier}
v(\tau, \bm{x}) = \int \frac {d^3k} {(2\pi)^3} \left[ v_{\bm{k}}(\tau)
{\rm e}^{i\bm{k}\cdot\bm{x}} + v^*_{\bm{k}}(\tau) {\rm e}^{-i\bm{k}\cdot\bm{x}}
\right]
\,.
\end{equation}
Note that $v$ is real by construction, whereas $v_{\bm{k}}$ is usually
complex. Moreover, $\bm{k}$ is defined in co--moving coordinates,
i.e. it remains unchanged by the expansion of the universe.

The MS equation in $\bm{k}-$space can be found by plugging
eq.(\ref{fourier}) into eq.(\ref{mukhanovdiff2})
\cite{Sasaki:1986hm, Mukhanov:1988jd, Mukhanov:1990me}:
\begin{equation} \label{mukhanov}
v_{\bm{k}}^{\prime\prime} + \left( k^2 - \frac {z^{\prime \prime}} {z} \right)
v_{\bm{k}} = 0\,.
\end{equation}
This equation describes how some perturbation with wave vector
$\bm{k}$ evolves with time. This equation has no general analytical
solution, due to the dependence on the background field dynamics via
$z^{\prime \prime}/z$. However, in some special cases, such as the SR
inflationary phase, an analytical solution exists, as we explain below.

\subsection{Quantization, Initial Condition and Bunch-Davies Vacuum}

Before discussing analytical solutions of eq.(\ref{mukhanov}) we first
describe the quantization of our field. After all, the physical origin
of the curvature perturbations generated by inflation are quantum
fluctuations of the inflaton field. Using the canonical quantization
procedure, we write the QFT analogue of the classical Fourier decomposition
of eq.(\ref{fourier}) \cite{Baumann:2009ds}:
\begin{equation} \label{qft1}
\hat{v} = \int \frac {d^3 k} {(2\pi)^3} \left[ v_{\bm{k}}(\tau) \hat{a}_{\bm{k}}
{\rm e}^{i\bm{k}\cdot\bm{x}}
+ v^{*}_{\bm{k}}(\tau) \hat{a}^{\dagger}_{\bm{k}} {\rm e}^{-i\bm{k}\cdot\bm{x}}
\right]\,,
\end{equation}
where $\hat a_{\bm{k}}$ and $\hat a^\dagger_{\bm{k}}$ are annihilation and
creation operators. The corresponding Fourier modes corresponding to a
fixed co--moving wave vector $\bm{k}$ are
\begin{equation} \label{qft2}
\hat{v}_{\bm{k}} = v_{\bm{k}}(\tau) \hat{a}_{\bm{k}}
+ v^{*}_{-\bm{k}}(\tau) \hat{a}^{\dagger}_{-\bm{k}}\,;
\end{equation}
note that $v_{\bm{k}}$ in eqs.(\ref{qft1}) and (\ref{qft2}) again satisfy the
(classical) MS equation (\ref{mukhanov}).

Similarly one can also introduce the quantum version of the canonical
momentum variable
$\pi = \frac {\partial {\cal L}_2} {\partial v'} = v'$:
\begin{equation} \label{piq}
\hat{\pi} = \int \frac {d^3 k} {(2\pi)^3} \left[ v_{\bm{k}}^{\prime}(\tau)
\hat{a}_{\bm{k}} {\rm e}^{i\bm{k}\cdot\bm{x}}
+ v^{*\prime}_{\bm{k}}(\tau) \hat{a}^{\dagger}_{\bm{k}}
{\rm e}^{-i\bm{k}\cdot\bm{x}}\right]\,.
\end{equation}
We impose the canonical commutation relation between $\hat{v}$ and its
conjugate momentum variable $\hat \pi$,
\begin{equation}
[ \hat{v}(\tau,\bm{x}),\, \hat{\pi}(\tau,\bm{y})] = i
\delta( \bm{x} -\bm{y})\,.
\end{equation}
From eqs.(\ref{qft1}) and (\ref{piq}) we see that this requires
\begin{equation}
\begin{split}
i \delta( \bm{x} - \bm{y}) &= \int \frac{d^3 k}{(2\pi)^3}
\frac {d^3 q} {(2\pi)^3} \Big[ v_{\bm{k}} v_{\bm{q}}^{\prime} \left(
\hat{a}_{\bm{k}} \hat{a}_{\bm{q}} - \hat{a}_{\bm{q}} \hat{a}_{\bm{k}} \right)
{\rm e}^{i(\bm{k}\cdot\bm{x} + \bm{q}\cdot\bm{y})} + v_{\bm{k}} v_{\bm{q}}^{\star \prime}
\left( \hat{a}_{\bm{k}} \hat{a}_{\bm{q}}^{\dagger} - \hat{a}^{\dagger}_{\bm{q}}
\hat{a}_{\bm{k}} \right) {\rm e}^{i(\bm{k}\cdot\bm{x} - \bm{q}\cdot\bm{y})}\\
&+ v_{\bm{k}}^{\star} v_{\bm{q}}^{ \prime} \left( \hat{a}^{\dagger}_{\bm{k}}
\hat{a}_{\bm{q}} - \hat{a}_{\bm{q}} \hat{a}^{\dagger}_{\bm{k}} \right)
{\rm e}^{i(\bm{q}\cdot\bm{y} - \bm{k}\cdot\bm{x})} +
v_{\bm{k}}^{\star \prime} v_{\bm{q}}^{ \star \prime} \left(
\hat{a}^{\dagger}_{\bm{k}} \hat{a}^{\dagger}_{\bm{q}} - \hat{a}^{\dagger}_{\bm{q}}
\hat{a}^{\dagger}_{\bm{k}} \right)
{\rm e}^{-i(\bm{k}\cdot\bm{x} + \bm{q}\cdot\bm{y})} \Big]\,.
\end{split}
\label{comutator}
\end{equation}
Eq.(\ref{comutator}) implies
\begin{equation}
-i (v_{\bm{k}} v^{*\prime}_{\bm{q}} - v^{\prime}_{\bm{k}} v^{*}_{\bf{q}})
[\hat{a}_{\bm{k}}, \hat{a}^{\dagger}_{\bm{q}}] = (2\pi)^3\delta(\bm{k}-\bm{q})\,;
\end{equation}
\begin{equation}
v_{\bm{k}} v_{\bm{q}}^{\prime} \left( \hat{a}_{\bm{k}} \hat{a}_{\bm{q}} -
\hat{a}_{\bm{q}} \hat{a}_{\bm{k}} \right) = 0\,;
\end{equation}
and
\begin{equation}
v_{\bf{k}}^{\star \prime} v_{\bm{q}}^{ \star \prime} \left(
\hat{a}^{\dagger}_{\bm{k}} \hat{a}^{\dagger}_{\bm{q}}
-\hat{a}^{\dagger}_{\bm{q}}\hat{a}^{\dagger}_{\bm{k}} \right) =0\,.
\end{equation}
Normalizing the mode functions such that
$-i (v_{\bm{k}} v^{*\prime}_{\bm{k}} - v^{\prime}_{\bm{k}}
v^{*}_{\bm{k}}) =1$ leads to the canonical commutation relations for
the annihilation and creation operators:
\begin{equation}
[\hat{a}_{\bm{k}},\hat{a}^{\dagger}_{\bm{q}}] = (2\pi)^3 \delta(\bm{k}-\bm{q})
\end{equation}
and
\begin{equation}
[\hat{a}_{\bm{k}}, \hat{a}_{\bm{q}}] =
[\hat{a}^{\dagger}_{\bm{k}}, \hat{a}^{\dagger}_{\bm{q}}] = 0\,.
\end{equation}
The vacuum state $|0\rangle$ is usually defined by
\begin{equation} \label{vac}
\hat{a}_{\bm{k}}|0\rangle = 0 \ \ \ \forall \bm{k}\,.
\end{equation}
Unfortunately this definition is not unique, since eq.(\ref{qft1})
only fixes the product $v_{\bm{k}} \hat a_{\bm{k}}$, i.e. the vacuum
state defined by eq.(\ref{vac}) depends on the form of mode function
$v_{\bf k}$. In other words, eq.(\ref{vac}) defines the vacuum state
uniquely only once $v_{\bm{k}}$ has been fixed.

To that end we consider the limit $\tau \to -\infty$, such that
$|k\tau| \gg 1$ or $k\gg aH$, where $k = |\bm{k}|$; this corresponds
to perturbations with wavelength much smaller than the Hubble
horizon. In this limit the mode function $v_{\bf{k}}$ behaves like a
massless field in Minkowski spacetime, since the $z$ term in
eq.(\ref{mukhanov}) can be neglected compared to $k^2$:
\begin{equation} \label{msk}
v^{\prime \prime}_{\bm{k}} + k^2v_{\bm{k}} = 0\,. 
\end{equation}
This describes a simple harmonic oscillator.

At this point we note that eq.(\ref{msk}), as well as the original MS
equation (\ref{mukhanov}), only depend on $k$. We can therefore make the
ansatz
\begin{equation} \label{prod}
v_{\bm{k}} = v_k \eta(\bm{k}/k)\,,
\end{equation}
where without loss of generality we can normalize the
angle--dependence $\eta$ such that $|\eta| = 1$, i.e. $\eta$ is a
time--independent pure phase, which factorizes in eqs.(\ref{mukhanov})
and (\ref{msk}). We then impose the boundary
condition\footnote{Formally this is an initial condition, although
	physically $\tau \rightarrow -\infty$ may well not fall into the
	inflationary epoch. Fortunately we saw in Sec.~3 that to good
	approximation this initial condition can be imposed at any time
	as long as the mode is still well within the horizon.}
\begin{equation} \label{initialcond}
\lim\limits_{\tau \to -\infty} v_k = \frac{e^{-ik\tau}}{\sqrt{2k}}.
\end{equation}
Eq.(\ref{initialcond}) fixes the mode function $v_k$ and thus also the
vacuum state (up to some angle--dependent phase factor, which is not
physically relevant); this is usually referred to as the Bunch--Davies
vacuum.

%%%%%%%%%%%%%%%%%%%%%%%%%%%%%%%%%%%%%%%%%%%%%%%%%%%%%%%%%%%%%

\subsection{Analytical Solution in  Quasi-de Sitter Spacetime}
\label{desitter}

We now describe the analytical solution of the MS equation in the
limit where the Hubble parameter $H$ is nearly constant.  This also
means that the Hubble SR parameter $\epsilon_H$ is small and nearly
constant, thus the time derivative of $\epsilon_H$ can be neglected;
these conditions are met during (U)SR inflation. Using eq.(\ref{za}),
we then obtain:
\begin{equation}
\frac {z^{\prime \prime}} {z} = \frac {a^{\prime \prime}} {a} = \frac {2} {\tau^2} \,.
\end{equation}
Inserting this into eq.(\ref{mukhanov}) yields
\begin{equation} \label{limitmu}
v^{\prime \prime}_k + \left( k^2 - \frac {2} {\tau^2} \right) v_k= 0 \,.
\end{equation}
The general analytical solution of this equation is given by
\cite{Baumann:2009ds}
\begin{equation}
v_k = \alpha \frac {{\rm e}^{-ik\tau}} {\sqrt{2k}} \left( 1
- \frac {i} {k\tau} \right) + \beta \frac {{\rm e}^{ik\tau}} {\sqrt{2k}}
\left( 1 + \frac {i} {k\tau} \right)\,,
\end{equation}
where $\alpha$ and $\beta$ are integration constants. The initial
conditions in eq.(\ref{initialcond}) imply $\alpha =1$ and $\beta =0$,
which leads to the Bunch--Davies mode functions
\begin{equation} \label{vkeq}
v_k = \frac {{\rm e}^{-ik\tau}} {\sqrt{2k}} \left( 1
- \frac {i} {k\tau} \right) \,.
\end{equation}

\subsection{Power Spectrum in Quasi-de Sitter Spacetime}

Having solved the Mukhanov--Sasaki equation, we can compute the power
spectrum of the field,
$\delta \hat{\phi}_{\bm{k}} \equiv a^{-1} \hat{v}_{\bm{k}}$:
\begin{equation} \label{powerspectrum}
\begin{split}
\langle 0| \delta \hat{\phi}_{\bm{k}}(\tau) \
\delta \hat{\phi}_{\bm{k}^{\prime}}(\tau) |0 \rangle &=
(2\pi)^3 \delta(\bm{k} + \bm{k}^{\prime}) \frac {\left| v_k(\tau) \right|^2}
{a^2} \\ & = (2\pi)^3 \delta(\bm{k} + \bm{k}^{\prime})
\frac {H^2} {2k^3} (1+k^2\tau^2)\,,
\end{split}
\end{equation}
where we have used eq.(\ref{vkeq}) as well as the expression for the
scale factor $a(\tau) = -\frac{1}{H\tau}$ which holds for constant
$H$, i.e. during (U)SR inflation. On super-horizon scales,
$|k\tau| \ll 1$ or equivalently $k \ll aH$, eq.(\ref{powerspectrum})
becomes
\begin{equation}
\langle 0| \delta \hat{\phi}_{\bm{k}}(\tau) \
\delta \hat{\phi}_{\bm{k}^{\prime}}(\tau) |0 \rangle
\to (2\pi)^3 \delta(\bm{k} + \bm{k}^{\prime}) \frac {H^2} {2k^3}\,,
\end{equation}
or in a dimensionless version (recall that we are using Planckian
units where $M_p =1$):
\begin{equation} \label{dimensionless}
\Delta^2_{\delta \phi } = \left(\frac{H}{2\pi}\right)^2.
\end{equation}
Eq.(\ref{dimensionless}) also implies
$\sqrt{ \langle \delta \phi_{\bm{k}}^2 \rangle} = H/(2\pi)$, which is
the frequently used formula for the quantum fluctuations of light
fields (with mass smaller than $H$) during SR inflation. As shown in
Section~\ref{srusr}, during SR inflation curvature perturbations are
frozen at super--horizon scale, thus the power spectrum can be
computed at the horizon crossing, i.e. for $a(t_k) H(t_k) = k$
\cite{Baumann:2009ds}:
\begin{equation}
\langle \zeta_{\bm{k}} \zeta_{\bm{k}^{\prime}} \rangle = (2\pi)^3
\delta(\bm{k} + \bm{k}^{\prime}) \frac {H^2_k H^2_k}
{2k^3 \dot{\phi}(t_k)^2}\,.
\end{equation}
The corresponding dimensionless power spectrum is
\begin{equation} \label{dimensionlesspower}
\mathcal{P}_{\zeta}(k) = \Delta_{\zeta}^2(k) = \frac{H^2_k H^2_k}
{ (2\pi)^2 \dot{\phi}(t_k)^2} = \frac {H^2_k} {8\pi^2 \epsilon_H},
\end{equation}
where we have used the definition
$\epsilon_H = \frac {1} {2} \left( \frac{\dot{\phi}} {H}
\right)^2$. Eq.(\ref{dimensionlesspower}) is widely used in the
literature when discussing SR inflation, where $\epsilon_H$
is small and its variation with time can be neglected. Moreover,
during SR inflation, the energy is mainly dominated by the
potential, thus we have $H^2 = \frac{V}{3}$. Using in addition the SR
solution for the equation of motion of the inflaton field,
$\dot{\phi} = \frac{-V^{\prime}}{3H}$, allows us to rewrite $\epsilon_H$
as:
\begin{equation} \label{power_H}
\epsilon_H = \frac {1} {2} \left( \frac {-V^{\prime}} {3H^2} \right)^2
= \frac {1} {2} \left( \frac {V^{\prime}} {V} \right)^2
\equiv \epsilon_V\,.
\end{equation}
$\epsilon_V$ is usually called the potential SR
parameter. This leads to another frequently used formula for
the power spectrum:
\begin{equation} \label{dimensionlesspower2}
\mathcal{P}_{\zeta} = \frac {V} {24 \ \epsilon_V \ \pi^2}\,.
\end{equation}
However, for non SR inflation -- in particular, during the
overshooting epoch which we have explored in this paper --
$\epsilon_H$ changes rapidly and $z^{\prime \prime}/z \neq
2/\tau^2$. There eqs.(\ref{dimensionlesspower}), (\ref{power_H}) and
eq.(\ref{dimensionlesspower2}) are no longer valid; note in particular
that eq.(\ref{dimensionlesspower2}) predicts a diverging power
spectrum at a true saddle point where $V^\prime$, and hence
$\epsilon_V$, vanishes. In this case we must solve the Mukhanov-Sasaki
eq.(\ref{mukhanov}) numerically in order to reliably estimate the
power spectrum at the end of inflation.

\end{document}